\documentclass[pre,twocolumn]{revtex4}

\usepackage{natbib}
\usepackage{amssymb,amsmath}
\usepackage{epsfig}
\usepackage{bm}

\usepackage{graphicx}

\usepackage{color}
\usepackage[normalem]{ulem}

\newcommand{\vicente}[1]{{ #1}}  
\newcommand{\wdf}[1]{{ #1}}      

\newcommand\beq{\begin{equation}}
\newcommand\eeq{\end{equation}}
\newcommand\beqa{\begin{eqnarray}}
\newcommand\eeqa{\end{eqnarray}}
\newcommand{\dd}{\text{d}}

\newcommand{\al}{\alpha}

\begin{document}

\title{Transport coefficients of solid particles immersed in a viscous gas}


\author{Vicente Garz\'{o}\footnote[1]{Electronic address: vicenteg@unex.es;
URL: http://www.unex.es/eweb/fisteor/vicente/}}
\affiliation{Departamento de F\'{\i}sica and Instituto de Computaci\'on Cient\'{\i}fica Avanzada (ICCAEx), Universidad de Extremadura, E-06071 Badajoz, Spain}
\author{William D. Fullmer\footnote[2]{Electronic address: william.fullmer@colorado.edu}}
\affiliation{Department of Chemical and Biological Engineering, University of Colorado, Boulder, CO 80309, USA}
\author{Christine M. Hrenya\footnote[3]{Electronic address: hrenya@colorado.edu}}
\affiliation{Department of Chemical and Biological Engineering, University of Colorado, Boulder, CO 80309, USA}
\author{Xiaolong Yin\footnote[4]{Electronic address: xyin@mines.edu}}
\affiliation{Petroleum Engineering Department, Colorado School of Mines, Golden, CO 80401, USA}

\begin{abstract}
Transport properties of a suspension of solid particles in a viscous gas are studied.  The dissipation in such systems arises from two sources: inelasticity in particle collisions and viscous dissipation due to the effect of the gas phase on the particles. Here, we consider a simplified case in which the mean relative velocity between the gas and solid phases is taken to be zero, such that ``thermal drag'' is the only remaining gas-solid interaction. Unlike the previous more general treatment of the drag force [Garz\'o \emph{et al.}, J. Fluid Mech. \textbf{712}, 129 (2012)], here we take into account contributions to the (scaled) transport coefficients $\eta^*$ (shear viscosity), $\kappa^*$ (thermal conductivity) and $\mu^*$ (Dufour-like coefficient) coming from the temperature-dependence of the (dimensionless) friction coefficient $\gamma^*$ characterizing the amplitude of the drag force. At moderate densities, the thermal drag model (which is based on the Enskog kinetic equation) is solved by means of the Chapman-Enskog method and the Navier-Stokes transport coefficients are determined in terms of the coefficient of restitution, the solid volume fraction and the friction coefficient. The results indicate that the effect of the gas phase on $\eta^*$ and $\mu^*$ is non-negligible (especially in the case of relatively dilute systems) while the form of $\kappa^*$ is the same as the one obtained in the dry granular limit. Finally, as an application of these results, a linear stability analysis of the hydrodynamic equations is carried out to analyze the conditions for stability of the homogeneous cooling state. A comparison with direct numerical simulations shows a good agreement for conditions of practical interest.
\end{abstract}



\draft
\date{\today}
\maketitle

\section{Introduction}
\label{sec1}

High-velocity, gas-solid flows occur in a wide range of practical applications like pneumatic transport lines and circulating fluidized beds, to name a few.  The earliest continuum descriptions of such multiphase flows (see, for example Ref.\ \cite{SJ89}) were based on their granular predecessors in which the role of the gas phase is negligible.  More specifically, an empirical drag law was added to the solids momentum balance, though the granular energy balance and solid-phase transport coefficients were used without any gas-phase modifications.  In the decades since those pioneering efforts, the effect of the gas phase on the granular energy balance (i.e., additional source/sink terms) and solid-phase transport coefficients has been incorporated with increasing rigor; see representative works in Refs.\ \cite{K90,G94,KS99} and a more detailed review in Ref.\ \cite{GTSH12}.

The most rigorous incorporation of gas-phase effects into solid-phase descriptions occurs at the starting point for the continuum derivation, namely the kinetic (Enskog) equation for the solids phase.  Numerous groups have taken such an approach by assuming that the instantaneous drag force appearing in the kinetic equation takes the same form as the mean drag force, except that instantaneous velocities are used in place of mean velocities (see Ref.\ \cite{GTSH12} for overview).  Recent work by Tenneti \emph{et al.} \cite{TGHFS10}, however, indicates that such an ad hoc assumption does not correctly capture the particle acceleration-velocity correlation observed via direct numerical simulations (DNS).  For the case of Stokes flow, the acceleration-velocity correlation has been correctly captured via analytical means \cite{K90}, though extensions beyond the low-Stokes limit are difficult due to inherent nonlinearities \cite[]{KH01}.  Moreover, previous analytical approaches have not accounted for the effects of the gas phase on transport coefficients but rather for the new sources/sinks of granular energy.

As an alternative to overcome past challenges with analytical approaches, Garz\'o  \emph{et al.} \cite{GTSH12} proposed an instantaneous particle acceleration based on a generalized Langevin model that was extracted from DNS simulations.  The model accounts for three sources of particle acceleration due to the gas phase: (i) mean drag (via a term involving a coefficient $\beta$ and mean gas/solid velocities), (ii) ``thermal'' drag (via a term involving a coefficient $\gamma$ and the granular temperature, or measure of particle velocity fluctuations), and (iii) particle neighbor effects (via a term involving a coefficient $\xi$ and stochastic Wiener process increment).  A benefit of using an acceleration model derived from DNS simulations is that in principle, it is not limited to extremes of the parameter space, unlike past analytical approaches. The new model was then incorporated into the starting Enskog kinetic equation to derive the balance equations and constitutive relations for gas-solid flows.  Attention was paid to the Stokes limit initially, in order to verify the correctness of the calculations, but more importantly to determine the effect of the gas phase on transport coefficients, which had not previously been done in a rigorous manner.  The results indicate that the effect of the gas phase on both the shear viscosity and the Dufour-like coefficients is non-negligible for industrially-relevant portions of the parameter space.

In the current effort, an extension of our previous study \cite{GTSH12} that more accurately accounts for the granular temperature dependence of the gas-phase effects on the transport coefficients is undertaken. In our previous work \cite{GTSH12}, the temperature dependence of two scaled parameters in the particle acceleration model, namely $\gamma^*$ and $\xi^*$, were implicitly neglected in order to obtain analytic (explicit) expressions for the transport coefficients.  Here, \vicente{the above temperature} dependence is considered for a simplified case, namely \vicente{when the mean relative velocities between the gas $\mathbf{U}_g$ and solid $\mathbf{U}$ phases is zero}, i.e.,  $\Delta {\bf U}\equiv {\bf U}-{\bf U}_g=\mathbf{0}$. More specifically, in such a simplified system only the thermal drag remains; i.e., mean drag and neighbor effects, which are proportional to $\Delta {\bf U}$ in the particle acceleration model, disappear.  This simplification is again motivated by a desire for analytic expressions when the full granular temperature dependence of $\gamma^*$ is considered.   It is worthwhile to note that the resulting (thermal) drag model, which is linear in granular temperature, has been recently employed \cite{WZLH09,H13,HT13,SMMD13,WGZS14,ChVG15} to model the effect of a viscous gas in gas-solid suspensions.

The outline of the paper is as follows. In Sec.\ \ref{sec2} the simplified model (thermal drag only) allowing for the temperature dependence of the (scaled) thermal drag coefficient $\gamma^*$ is introduced and the corresponding balance equations for the densities of mass, momentum and energy are derived. Section \ref{sec3} deals with the so-called homogeneous cooling state (HCS) where a scaling solution is proposed that depends on granular temperature $T$ only through the dimensionless velocity $\mathbf{c}=\mathbf{v}/v_0(T)$, where $v_0(T)=\sqrt{2T/m}$ is the thermal velocity. This solution is similar to the one obtained before in previous works on \emph{dry} granular gases \cite{NE98}. The Chapman-Enskog method is carried out in Sec.\ \ref{sec4} to solve the Enskog kinetic equation up to first order in the spatial gradients, i.e., Navier-Stokes order. In Sec.\ \ref{sec5}, the resulting transport coefficients are obtained  and then compared to the previous treatment \cite{GTSH12} where the temperature dependence of the (scaled) thermal drag coefficient $\gamma^*$ was neglected but $\Delta {\bf U}\neq \mathbf{0}$ was allowed.  With the exception of the shear viscosity and Dufour-like coefficient for quite dissipative particles (coefficient of restitution $\alpha \lesssim 0.7$) and high values of $\gamma^*$, the transport coefficients derived here match well with prior results \cite{GTSH12}, providing confidence in the approximate, but more general expressions for which $\Delta {\bf U}\neq \mathbf{0}$.  Then in Sec.\ \ref{sec6}, as an application of the new theory, a linear stability analysis  is performed using the HCS as the basis state. Section \ref{sec7} provides some technical details on the DNS performed here while the theoretical results derived from the stability analysis are compared with DNS results in Sec.\ \ref{sec8} for conditions of practical interest.  Good agreement is found, providing quantitative validation of the calculations carried out here. Finally, the paper is closed in section \ref{sec9} with some concluding remarks.

\section{Thermal drag model for gas-solid flows}
\label{sec2}


We consider a system of solid particles suspended in a viscous gas. Under rapid flow conditions, particles are usually modelled as a gas of smooth \emph{inelastic} hard spheres or disks. In this case, the inelasticity of collisions is characterized by a \emph{constant} (positive) coefficient of normal restitution $\al \leq 1$. The case $\al=1$ corresponds to elastic collisions. The suspension is in the presence of the gravitational force $m \mathbf{g}$, where $m$ is the particle mass and $\mathbf{g}$ is the gravity acceleration. For moderate densities, it is assumed that the one-particle velocity distribution function $f(\mathbf{r}, \mathbf{v};t)$ is accurately described by the Enskog kinetic equation \cite{BDS97,BP04}
\beqa
\partial_{t}f + \mathbf{v}\cdot \nabla f + \frac{\partial}{\partial \mathbf{v}} \cdot
\left[\left(\frac{\mathbf{F}_{\text{fluid}}}{m} \right) f \right] +
\mathbf{g} \cdot \frac{\partial f}{\partial \mathbf{v} } = J\left[f,f\right],\nonumber\\
\label{2.0a}
\eeqa
where $\mathbf{v}$ is the particle velocity and $\mathbf{F}_{\text{fluid}}$ denotes the fluid-solid interaction force that models the effect of the viscous gas on solid particles. In order to fully account for the influence of the interstitial molecular fluid on the dynamics of grains, a instantaneous fluid force model  has been recently proposed \cite{GTSH12}. In this model, the instantaneous impulse is given by
\beq
m \dd \mathbf{v} =  {\mathbf F}_{\text{fluid}} \dd t=
- \beta \Delta \mathbf{U} \dd t - \gamma \mathbf{V} \dd t + m \sqrt {\xi}  \dd \mathbf{W},
\label{2.0b}
\eeq
where $\mathbf{V}=\mathbf{v}-\mathbf{U}$ is the particle fluctuation (or peculiar) velocity, the vector $\dd \mathbf{W}$ is a Wiener process increment (stochastic term) and the scalars $\beta$, $\gamma$ and $\xi$ are the model coefficients \cite{GTSH12}. According to Eq.\ \eqref{2.0b}, the fluid-solid interaction force present in high-velocity gas-solid flows is constituted by three different terms: (i) a term proportional to the difference between the mean velocities of gas and solid phases $\Delta \mathbf{U}$ (mean drag), (ii) an additional contribution to the drag force arising from particle velocity fluctuations (thermal drag) and (iii) a stochastic Langevin-like term that accounts for added effects of neighboring particles (neighbor effect).
Decomposing the deterministic fluid force into mean and fluctuating components allows for the distinct mean ($\beta$) and thermal ($\gamma$) drag models. DNS data has suggested that beyond the dilute limit, the concentration dependence of these two models is no longer identical \cite{KS99}. The first and  third terms are proportional to $\Delta \mathbf{U}$ and hence, they vanish for particular situations where the mean velocity of solid particles follows the mean flow velocity of fluid (like for instance in the simple shear flow state \cite{ChVG15,TK95,SMTK96}).

For the most general condition of $\Delta {\bf U}\neq \mathbf{0}$, the kinetic equation for gas-solid suspensions is given by \cite{GTSH12}
\beqa
\partial_{t}f&+&\mathbf{v}\cdot \mathbf{\nabla}f-\frac{\beta}{m}\Delta {\bf U}\cdot
\frac{\partial f}{\partial {\bf V}}-\frac{\gamma}{m} \frac{\partial}{\partial
{\bf V}}\cdot {\bf V} f\nonumber\\
& & -\frac{1}{2}\xi\frac{\partial^2 f}{\partial V^2}+
\mathbf{g} \cdot \frac{\partial f}{\partial {\bf V}}=J_{\text{E}}\left[{\bf r}, {\bf v}|f,f\right], \label{2.1}
\eeqa
where
\beqa
\label{2.2}
J_{\text{E}}\left[{\bf r}, {\bf v}_{1}|f,f\right] &=&\sigma^{d-1}\int \dd{\bf v}
_{2}\int \dd\widehat{\boldsymbol{\sigma}}\,\Theta (\widehat{{\boldsymbol {\sigma }}}
\cdot {\bf g}_{12})(\widehat{\boldsymbol {\sigma }}\cdot {\bf g}_{12})  \nonumber\\
&  & \times \left[ \alpha^{-2}\chi({\bf r},{\bf r}-\boldsymbol {\sigma })
f({\bf r}, {\bf v}_1';t)\right. \nonumber\\
& & \times f({\bf r}-\boldsymbol {\sigma}, {\bf v}_2';t)-
\chi({\bf r},{\bf r}+\boldsymbol {\sigma })
\nonumber\\
& &\times \left.
f({\bf r}, {\bf v}_1;t) f({\bf r}+\boldsymbol {\sigma }, {\bf v}_2;t)\right]
\eeqa
is the Enskog collision operator. Like the Boltzmann equation, the Enskog equation neglects velocity correlations among particles that are about to collide, but it takes into account the dominant spatial correlations due to excluded-volume effects.

According to Eq.\ \eqref{2.1}, gas-phase effects appear in terms involving $\beta$ (mean drag), $\gamma$ (thermal drag) and $\xi$ (neighbor effects). In Eq.\ \eqref{2.2}, $d$ is the dimensionality of the system ($d=2$ for disks and $d=3$ for spheres), $\boldsymbol {\sigma}=\sigma \widehat{\boldsymbol {\sigma}}$,  $\widehat{\boldsymbol {\sigma}}$ being a unit vector and $\sigma$ the particle diameter, $\Theta$ is the Heaviside step function, ${\bf g}_{12}={\bf v}_{1}-{\bf v}_{2}$ and $\chi[{\bf r},{\bf r}+\boldsymbol {\sigma}|\{n(t)] $ is the equilibrium pair correlation function at contact as a functional of the nonequilibrium density field $n({\bf r}, t)$ defined by
\beq
\label{2.2.1}
n({\bf r}, t)=\int\; \dd{\bf v} f({\bf r},{\bf v},t).
\eeq
For the case of spheres ($d=3$) considered in this work, the Carnahan-Starling approximation \cite{CS69} for $\chi$ is given by
\begin{equation}
\label{2.2.2}
\chi(\phi)=\frac{1-\frac{1}{2}\phi}{(1-\phi)^3},
\end{equation}
where
\begin{equation}
\label{2.7}
\phi=\frac{\pi ^{d/2}}{2^{d-1}d\Gamma \left(\frac{d}{2}\right)}n\sigma^d
\end{equation}
is the solid volume fraction. The primes on the velocities in Eq.\ \eqref{2.2} denote the initial values $\{{\bf v}_{1}^{\prime},
{\bf v}_{2}^{\prime }\}$ that lead to $\{{\bf v}_{1},{\bf v}_{2}\}$
following a binary collision:
\begin{subequations}
\begin{equation}
{\bf v}_{1}^{\prime}={\bf v}_{1}-\frac{1}{2}\left( 1+\alpha^{-1}\right)
(\widehat{{\boldsymbol {\sigma }}}\cdot {\bf g}_{12})\widehat{{\boldsymbol {\sigma }}},
\eeq
\beq
{\bf v}_{2}^{\prime }={\bf v}_{2}+\frac{1}{2}\left( 1+\alpha^{-1}\right)
(\widehat{{\boldsymbol {\sigma }}}\cdot {\bf g}_{12})\widehat{
\boldsymbol {\sigma}}. \label{2.3}
\end{equation}
\end{subequations}
Moreover, in Eq.\ \eqref{2.1},  we recall that $\Delta {\bf U}={\bf U}-{\bf U}_g$ where $\mathbf{U}_g$ is the (known) mean flow velocity of the gas phase and
\beq
\label{2.4}
\mathbf{U}=\frac{1}{n(\mathbf{r},t)}\int\; \dd \mathbf{v}\; \mathbf{v} f(\mathbf{r},\mathbf{v},t)
\eeq
is the mean particle velocity. The model coefficients $\beta$, $\gamma$ and $\xi$ are extracted from DNS simulations \cite{GTSH12}. Accordingly, these coefficients depend on constant parameters (particle mass and diameter, gas viscosity) as well as on the hydrodynamic (mean) variables (solids concentration, gas and solid velocities, and granular temperature). In particular, in the case of low mean flow Reynolds numbers, $Re_m = (1-\phi) \sigma \rho_g |\Delta {\bf U}| / \mu_g$, the expressions of $\gamma$ and $\xi$ for hard spheres ($d=3$) are
\begin{equation}
\label{2.5}
\gamma=\frac{m}{\tau_g}R_\text{diss}(\phi),
\end{equation}
\begin{equation}
\label{2.6}
\xi=\frac{1}{6\sqrt{\pi}}\frac{\sigma |\Delta {\bf U}|^2}{\tau_g^2
\sqrt{\frac{T}{m}}}S^*(\phi),
\end{equation}
where $\tau_g=m/(3\pi \mu_g \sigma)$ is the viscous relaxation time, $\mu_g$ is the gas-phase dynamic (shear) viscosity and
\beq
\label{2.6.1}
T(\mathbf{r},t)=\frac{2}{d n(\mathbf{r},t)}\int\; \dd \mathbf{v}\; V^2 \; f(\mathbf{r},\mathbf{v},t)
\eeq
is the \emph{granular} temperature. In Eqs.\ \eqref{2.5} and \eqref{2.6}, $R_\text{diss}(\phi)$ and $S^*(\phi)$ are only functions of the solid volume fraction $\phi$. Approximate forms for $R_\text{diss}(\phi)$ (see Eq.\ \eqref{5.19}) and $S^*(\phi)$ can be found in the literature \cite{K90,KS99,WKL03}.

It is quite apparent that in the suspension kinetic model defined by Eqs.\ \eqref{2.1} and \eqref{2.2}, the form of the Enskog collision operator $J_\text{E}[\mathbf{r}, \mathbf{v}|f,f ]$ is
the same as for a dry granular gas and hence, the collision
dynamics do not contain any effects related to the interstitial fluid. As has
been previously discussed in several papers \cite{K90,TK95,SMTK96,WKL03}, the
above assumption requires that the mean-free time between
collisions is much less than the time taken by the fluid forces to significantly affect the motion of
solid particles (i.e., the viscous relaxation time). Under these conditions, it is expected that the suspension model defined by Eq.\ \eqref{2.1} will accurately describe situations where the stresses exerted by the interstitial
fluid on particles are sufficiently small so that they only have a weak
influence on the dynamics of grains. As the particle-to-fluid density ratio decreases
(e.g., glass beads in liquid water), the above assumption may not be reliable and hence, one may need to consider the effect of the interstitial fluid on the collision operator.

The macroscopic balance equations for the densities of mass, momentum and energy can be exactly derived from the Enskog equation \eqref{2.1}. The are given by \cite{GTSH12}
\begin{equation}
D_{t}n+n\nabla \cdot {\bf U}=0\;, \label{2.8}
\end{equation}
\begin{equation}
D_{t}{\bf U}+\rho ^{-1}\nabla \cdot {\sf P}=-\frac{\beta}{m}\Delta {\bf U}+{\bf g}\;,
\label{2.9}
\end{equation}
\begin{equation}
D_{t}T+\frac{2}{dn} \left( \nabla \cdot {\bf q}+{\sf P}:\nabla {\bf U}\right) =
-\frac{2 T}{m}\gamma +m \xi-\zeta \,T\;.
\label{2.10}
\end{equation}
In the above equations, $D_{t}=\partial_{t}+{\bf U}\cdot \nabla$ is the material
derivative and $\rho = m n \equiv \rho_s \phi$ is the mass density ($\rho_s$ is the material density of a particle). The presence of the gas phase gives rise to three new terms: mean drag (first term on the right hand side of Eq.\ \eqref{2.9}), thermal drag (first term on the right hand side of Eq.\ \eqref{2.10}), and associated neighbor effects (second term on the right hand side of Eq.\ \eqref{2.10}). In addition, the cooling rate $\zeta$ is proportional to $1-\alpha^2$ and is due to dissipative collisions. The pressure tensor ${\sf P}({\bf r},t)$ and the heat flux ${\bf q}({\bf r},t)$ have both {\em kinetic} and {\em collisional transfer} contributions, i.e., ${\sf P}={\sf P}^k+{\sf P}^c$ and ${\bf q}={\bf q}^k+{\bf q}^c$. The kinetic contributions are given by
\begin{equation}
\label{2.11}
{\sf P}^k=\int \; \dd{\bf v} m{\bf V}{\bf V}f({\bf r},{\bf v},t),
\end{equation}
\begin{equation}
\label{2.11.1}
{\bf q}^k=\int \; \dd{\bf v} \frac{m}{2}V^2{\bf V}f({\bf r},{\bf v},t),
\end{equation}
and the definition of the collisional transfer contributions ${\sf P}^c$ and $\mathbf{q}^c$ are given by Eqs.\ (4.11) and (4.12), respectively, of Ref.\ \cite{GTSH12}. \vicente{Since the forms of the collisional contributions to ${\sf P}^c$ and $\mathbf{q}^c$ are not affected by the inclusion of the temperature dependence of $\gamma^*$, their Navier-Stokes  expressions (first order in spatial gradients) are the same as those derived before in Ref.\ \cite{GTSH12}.} The cooling rate is given by
\beqa
\zeta &=&\frac{\left(1-\alpha^{2}\right)}{4dnT} m \sigma^{d-1}\int \dd\mathbf{v}
_{1}\int \dd\mathbf{v}_{2}\int \dd\widehat{\boldsymbol {\sigma }}
\Theta (\widehat{\boldsymbol {\sigma }}\cdot
\mathbf{g}_{12})\nonumber\\
&\times& (\widehat{ \boldsymbol {\sigma }}\cdot
\mathbf{g}_{12})^{3}
f^{(2)}(\mathbf{r}, \mathbf{r}+\boldsymbol {\sigma},\mathbf{v}_{1},\mathbf{v}_{2};t), \label{2.12}
\eeqa
where
\beq
\label{2.12.1}
f^{(2)}(\mathbf{r}_1, \mathbf{r}_2, \mathbf{v}_1, \mathbf{v}_2,t)=\chi(\mathbf{r}_1,\mathbf{r}_2|n(t))f(\mathbf{r}_1,\mathbf{v}_1,t)f(\mathbf{r}_2,\mathbf{v}_2,t).
\eeq

Needless to say, the hydrodynamic balance equations \eqref{2.8}--\eqref{2.10} are not a \emph{closed} set of equations for the hydrodynamic fields unless the pressure tensor, the heat flux and the cooling rate are expressed as functionals of the fields $n$, $\mathbf{U}$, and $T$. This task can be accomplished by solving the corresponding kinetic equation by means of the Chapman-Enskog method \cite{CC70}. This perturbation method was used in Ref.\ \cite{GTSH12} to determine the pressure tensor $P_{ij}^{(1)}$ and the heat flux $\mathbf{q}^{(1)}$ to first-order in spatial gradients. Their expressions are
\begin{equation}
\label{5.1}
P_{ij}^{(1)}=-\eta\left( \partial_{i}U_{j}+\partial _{j
}U_{i}-\frac{2}{d}\delta _{ij}\nabla \cdot
\mathbf{U} \right) -\lambda  \nabla \cdot
\mathbf{U},
\end{equation}
\beq
\label{5.2}
\mathbf{q}^{(1)}=-\kappa \nabla T-\mu \nabla n,
\eeq
where $\eta$ is the shear viscosity, $\lambda$ is the bulk viscosity, $\kappa$ is the thermal conductivity coefficient, and $\mu$ is a Dufour-like coefficient. While $\eta$, $\kappa$ and $\mu$ have kinetic and collisional contributions, $\lambda$ has only a collisional contribution. The Navier-Stokes transport coefficients can be written as
\beq
\label{5.3}
\eta\equiv \eta_0 \eta^*, \quad \lambda\equiv \eta_0 \lambda^*,
\quad \kappa\equiv \kappa_0 \kappa^*, \quad \mu\equiv \frac{T\kappa_0}{n}\mu^*,
\eeq
where $\eta_0=nT/\nu(T)$ is the shear viscosity of a molecular (dry) dilute gas and
\beq
\label{5.4}
\kappa_0=\frac{d(d+2)}{2(d-1)}\frac{\eta_0}{m}
\eeq
is the thermal conductivity of a molecular (dry) dilute gas. In addition,
\beq
\label{3.6}
\nu(T)=\frac{8}{d+2}\frac{\pi^{\left(d-1\right) /2}}{\Gamma \left( \frac{d}{2}\right)}
n\sigma^{d-1}\sqrt{\frac{T}{m}}
\eeq
is the collision frequency associated with the shear viscosity of a dilute elastic gas.

The (scaled) transport coefficients $\eta^*$, $\lambda^*$, $\kappa^*$ and $\mu^*$ are nonlinear functions of the solid volume fraction $\phi$, the coefficient of restitution $\al$, and the (dimensionless) coefficients
\beq
\label{2.13}
\gamma^*\equiv \frac{\gamma}{m\nu(T)}, \quad \xi^*\equiv \frac{m\xi}{T \nu(T)}.
\eeq
For the sake of simplicity, the expressions of the kinetic contributions to $\eta^*$, $\kappa^*$ and $\mu^*$ were derived in Ref.\ \cite{GTSH12} by neglecting the temperature-dependence of $\gamma^*$ and $\xi^*$. Thus, a natural question is whether, and if so to what extent, the conclusions drawn before \cite{GTSH12} may be altered when the above new ingredient is accounted for in the theory.

Nevertheless, the determination of the Navier-Stokes transport coefficients by considering the dependence of $\gamma^*$ and $\xi^*$ on $T$ by starting from the (complete) Langevin-like model \eqref{2.1} is very complex, especially if one wants to provide explicit expressions for the above coefficients. Thus, in order to gain some insight into the general problem, we consider here a simplified version of the model \eqref{2.1} where the mean flow velocities of solid particles and gas phase are assumed to coincide ($\Delta \mathbf{U} = 0$) \vicente{and hence according to Eq.\ \eqref{2.6}}, $\xi = 0$. In other words, the mean drag and neighbor effects are assumed to be negligible. As we will show below, the use of this simplified model allows one to get analytical results for the transport coefficients for general \emph{unsteady} conditions.

Therefore, in the case $\Delta \mathbf{U}=0$, the kinetic equation \eqref{2.1} reads
\begin{equation}
\partial_{t}f+\mathbf{v}\cdot \mathbf{\nabla}f-\frac{\gamma}{m} \frac{\partial}{\partial
{\bf V}}\cdot {\bf V} f+
\mathbf{g} \cdot \frac{\partial f}{\partial {\bf V}}=J_{\text{E}}\left[{\bf r}, {\bf v}|f,f\right],
\label{2.14}
\end{equation}
while the momentum and energy balance equations \eqref{2.9} and \eqref{2.10} become, respectively
\begin{equation}
D_{t}{\bf U}+\rho ^{-1}\nabla \cdot {\sf P}={\bf g}\;,
\label{2.15}
\end{equation}
\begin{equation}
D_{t}T+\frac{2}{dn} \left( \nabla \cdot {\bf q}+{\sf P}:\nabla {\bf U}\right) =
-\frac{2 T}{m}\gamma -\zeta \,T\;.
\label{2.16}
\end{equation}
The objective now is to solve the simplified kinetic equation \eqref{2.14} for states close to the HCS. \vicente{As mentioned in the Introduction}, it must be remarked that the same kinetic equation \eqref{2.14} has been previously used to study simple shear flows in gas-solid suspensions \cite{ChVG15,TK95,SMTK96}, particle clustering due to hydrodynamic interactions \cite{WK00}, steady states of particle systems driven by a vibrating boundary \cite{WZLH09} and more recently \cite{H13,HT13,SMMD13,WGZS14} to analyze the shear rheology of frictional hard sphere suspensions.

\section{Homogeneous cooling state}
\label{sec3}

The HCS is an ideal first test of the simplified kinetic equation since the mean motion of each phase (gas and solids) is zero, and thus $\Delta \mathbf{U}=\mathbf{0}$. In this case, we consider an isolated gas ($\mathbf{g}=\mathbf{0}$) where the density $n$ is constant and the time-dependent temperature $T(t)$ is spatially uniform. Consequently, the Enskog equation \eqref{2.14} for the homogeneous distribution $f_h$ becomes
\begin{equation}
\label{3.1}
\frac{\partial f_h}{\partial t}-\frac{\gamma}{m} \frac{\partial}{\partial
{\bf v}}\cdot {\bf v} f_h=\chi J_\text{B}[f_h,f_h],
\end{equation}
where here $J_\text{B}[f_h,f_h]$ is the Boltzmann collision operator for inelastic collisions, namely,
\beqa
\label{3.2}
J_{\text{B}}\left[f,f\right]&=&\sigma^{d-1}\int \dd{\bf v}
_{2}\int \dd\widehat{\boldsymbol{\sigma}}\,\Theta (\widehat{{\boldsymbol {\sigma }}}
\cdot {\bf g}_{12})(\widehat{\boldsymbol {\sigma }}\cdot {\bf g}_{12})\nonumber\\
& & \times \left[\alpha^{-2}
f({\bf v}_1')f({\bf v}_2')- f({\bf v}_1)f({\bf v}_2)\right].
\eeqa
The balance equations for the HCS yield $\partial_t n=0$, $\partial_t \mathbf{U}=\mathbf{0}$ and
\beq
\label{3.3}
\partial_t T=-\left(\zeta+\frac{2}{m}\gamma\right) T.
\eeq
Upon deriving Eq.\ \eqref{3.3} we have accounted for that the heat flux vanishes and the pressure tensor is diagonal, namely, $P_{ij}=p\delta_{ij}$ where \cite{GTSH12}
\begin{equation}
\label{3.4}
p=nT\left[1+2^{d-2}(1+\alpha)\chi \phi\right],
\end{equation}
is the hydrostatic pressure. Note that the presence of the gas phase does not enter in the constitutive relation for pressure. The solution to Eq.\ \eqref{3.3} can be written as \cite{YZMH13}
\beq
\label{3.5}
\frac{T(t)}{T_0}=\frac{4\gamma_0^{*2}e^{-2\gamma_0^*t^*}}{\left[2\gamma_0^*+
\zeta^*\left(1-e^{-\gamma_0^*t^*}\right)\right]^2}.
\eeq
Here, $T_0\equiv T(0)$ is the initial temperature, $\gamma_0^*\equiv \gamma/(m\nu(T_0))$, and $\zeta^*\equiv \zeta/\nu(T)$ where $\nu(T)$ is defined by Eq.\ \eqref{3.6}. Moreover, in Eq.\ \eqref{3.5}, $t^*\equiv \nu(T_0)t$. In order to get the explicit dependence of $T(t)/T_0$ on the coefficient of restitution $\al$ and the friction coefficient $\gamma$, one has to determine the (reduced) cooling rate $\zeta^*$.

In the hydrodynamic regime, since the time dependence of $f_h$ only occurs through the granular temperature $T$, then
\beq
\label{3.7}
\frac{\partial f_h}{\partial t}=
\frac{\partial f_h}{\partial T}\frac{\partial T}{\partial t}=-\left(\zeta+\frac{2}{m}\gamma\right)T\frac{\partial f_h}{\partial T},
\eeq
and Eq.\ \eqref{3.1} becomes
\beq
\label{3.8}
-\left(\zeta+\frac{2}{m}\gamma\right)T\frac{\partial f_h}{\partial T}-\frac{\gamma}{m} \frac{\partial}{\partial
{\bf v}}\cdot {\bf v} f_h=\chi J_\text{B}[\mathbf{v}|f_h,f_h].
\eeq
In the absence of the viscous drag force ($\gamma=0$), Eq.\  \eqref{3.8} admits the solution \cite{NE98}
\beq
\label{3.9}
f_h(\mathbf{v})=n v_0^{-d} \varphi_h (\mathbf{c}),
\eeq
where the scaling distribution $\varphi_h$ is an unknown function of the dimensionless velocity
\beq
\label{3.10}
\mathbf{c}=\frac{\mathbf{v}}{v_0},
\eeq
where $v_0=\sqrt{2T/m}$ is the thermal velocity. When $\gamma \neq 0$, according to the previous results derived for driven granular gases \cite{GMT12,GMV13,MVG13}, the scaled distribution $\varphi_h$ could have an additional dependence on the granular temperature through the dimensionless friction coefficient $\gamma^*=\gamma/(m\nu(T))$. On the other hand, it can be seen by direct substitution that the form \eqref{3.9} is still a solution of Eq.\ \eqref{3.8} and hence $\varphi_h$ does not explicitly depend on $\gamma^*$. This conclusion is consistent with the results obtained in Ref.\ \cite{L01} where it has been shown that the drag force term $\partial_\mathbf{v} \cdot \mathbf{v} f$ arises from a logarithmic change in the time-scale of the hard sphere system without external force.

Thus, according to the scaling \eqref{3.9}, one has the property
\beq
\label{3.11}
T\frac{\partial f_h}{\partial T}=-\frac{1}{2}\frac{\partial}{\partial
{\bf v}}\cdot {\bf v} f_h,
\eeq
and Eq.\ \eqref{3.8} reduces to
\beq
\label{3.12}
\frac{1}{2}\zeta \frac{\partial}{\partial{\bf v}}\cdot {\bf v} f_h=\chi J_\text{B}[f_h,f_h].
\eeq
Equation \eqref{3.12} is fully equivalent to the one obtained in the HCS of a dry granular gas (namely, when $\gamma^*=0$).

\begin{figure}
\includegraphics[width=0.85\columnwidth]{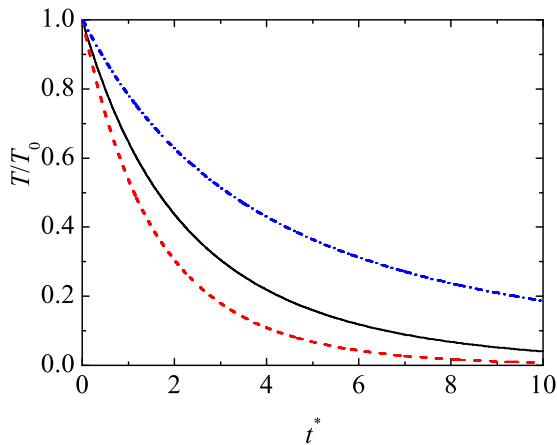}
\centering
\caption{(Color online) Temperature versus (dimensionless) time $t^*$ for a three-dimensional system with $\phi=0.2$ and $\al=0.8$. Three different values of the initial dimensionless friction coefficient $\gamma_0^*$ have been considered: $\gamma_0^*=0.2$ (dashed red line), $\gamma_0^*=0.1$ (solid line), and the dry granular case $\gamma_0^*=0$ (blue dash-dotted line).
\label{fig0}}
\end{figure}

To confirm the scaling \eqref{3.9}, let us analyze the evolution of the kurtosis or fourth-cumulant
\begin{equation}
\label{3.13}
a_{2}=\frac{1}{d(d+2)}\frac{m^2}{nT^2}\int\; \dd{\bf v}\; v^4 f_h(\mathbf{v})-1.
\end{equation}
Although the exact form of the homogeneous distribution function is not known, the knowledge of $a_2$ provides an indirect information of the deviation of $\varphi_h$ from its Gaussian form. In order to determine $a_2(t)$, we multiply Eq.\ \eqref{3.1} by $v^4$ and integrate over velocity. Neglecting nonlinear terms in $a_2$, the result can be written as \cite{NE98}
\beq
\label{3.14}
\frac{\partial a_2}{\partial \tau}+X a_2=Y,
\eeq
where
\beq
\label{3.15}
X=\frac{1+\al}{64d}\left[9+24d-\al(41-8d)+30\al^2(1-\al)\right]\chi,
\eeq
\beq
\label{3.16}
Y=\frac{(1-\al^2)(1-2\al^2)}{4d}\chi,
\eeq
and we have introduced the dimensionless time scale
\beq
\label{3.18}
\tau(t)=\int_0^t\; \dd t' \nu(t').
\eeq
The parameter $\tau$ is therefore an average number of collisions per particle in the time interval between $0$ and $t$. The solution to Eq.\ \eqref{3.14} is
\beq
\label{3.19}
a_2(\tau)=a_2(0)\;e^{-X \tau}+a_{2,\text{dry}},
\eeq
where $a_2(0)$ denotes the initial value of $a_2$ and
\beq
\label{3.20}
a_{2,\text{dry}}=\frac{16(1-\al)(1-2\al^2)}{9+24d-\al(41-8d)+30\al^2(1-\al)}
\eeq
is the value of $a_2$ in the case of a \emph{dry} granular gas \cite{NE98}. Since  $X>0$ in the entire range of values of $\al$, then $a_2\to a_{2,\text{dry}}$ and the results obtained for the (scaled) fourth-degree moment of $f_h$ in the presence or in the absence of the drag force are the same. To first order in $a_2$, the collisional cooling rate $\zeta$ can be written as
\beq
\label{3.21}
\zeta=\frac{d+2}{4d}(1-\al^2)\chi \left(1+\frac{3}{16}a_{2,\text{dry}}\right)\nu.
\eeq

Once the cooling rate is known, it is interesting to write the solution of Eq.\ \eqref{3.3} in terms of the (dimensionless) time $\tau$. The physical solution compatible with the dry granular limit case (no gas phase) is given by
\beq
\label{3.21.1}
\frac{T(\tau)}{T_0}=\frac{\left[2\gamma_0^*-e^{-\zeta^*\tau}\left(2\gamma_0^*+\zeta^*\right)\right]^2}{\zeta^{*2}}.
\eeq
In the case of elastic collisions ($\zeta^*=0$), Eq.\ \eqref{3.21.1} leads to $T(\tau)=T_0\left(1-2\gamma_0^*\tau \right)^2$ while in the absence of the gas phase ($\gamma_0^*=0$), Eq.\ \eqref{3.21.1} yields $T(\tau)=T_0 \exp(-2\zeta^*\tau)$. The latter expression is consistent with the results derived for granular gases \cite{NE98}. The relationship between the real time $t^*$ and the (reduced) time variable $\tau$ can be easily obtained from the identity
\beq
\label{3.21.2}
\tau=\frac{1}{2}\int_0^t\; \nu(T(t')) \dd t'=\frac{1}{2}\int_0^{t^*}\; \sqrt{\frac{T(t^*)}{T_0}}\dd t^*.
\eeq
The integration of Eq.\ \eqref{3.21.2} gives
\beq
\label{3.21.3}
\tau=\frac{\ln\left[-\zeta^*+e^{\gamma_0^* t^*}\left(2\gamma_0^*+\zeta^*\right)\right]-\gamma_0^* t^*-\ln(2\gamma_0^*)}
{\zeta^*}.
\eeq
Note that when $t^*\to \infty$, then $\tau \to \tau_0$ where
\beq
\label{3.21.4}
\tau_0=\frac{1}{\zeta^*}\ln \left( 1+\frac{\zeta^*}{2\gamma_0^*}\right).
\eeq

Figure \ref{fig0} illustrates the time dependence of the temperature for hard spheres ($d=3$) with $\phi=0.2$ and $\al=0.8$. Three values of $\gamma_0^*$ have been considered, including the dry granular limit case ($\gamma_0^*=0$). As expected, the temperature decays in time more slowly in the dry limit case than in the case of viscous suspensions. In addition, this decay is more pronounced as the effect of viscous gas increases. It must be remarked that the analytical result \eqref{3.5} compares quite well (before the onset of vortex instability) with DNS results \cite{YZMH13} in conditions of practical interest.

\section{Chapman-Enskog method}
\label{sec4}

We assume now that we slightly perturb the HCS analyzed in Sec.\ \ref{sec3} by small spatial gradients. In this case, there are non-diagonal contributions to the pressure tensor, the heat flux is different from zero and hence, one can identify the relevant Navier-Stokes transport coefficients of the suspension. The evaluation of these coefficients as functions of both the coefficient of restitution $\al$ and the friction coefficient $\gamma$ is the main goal of this paper.

Since the strength of the spatial gradients is small, the Enskog equation \eqref{2.14} is solved by means of the Chapman-Enskog method \cite{CC70} adapted to dissipative dynamics. The Chapman-Enskog method assumes the existence of a normal solution such that all space and time dependence of the distribution function occurs through the hydrodynamic fields
\begin{equation}
f({\bf r},{\bf v},t)=f\left[{\bf v}|n ({\bf r}, t),
T({\bf r}, t), {\bf U}({\bf r}, t) \right] \;.
\label{4.1}
\end{equation}
The notation on the right hand side indicates a functional dependence on the density, temperature and flow velocity. For small spatial variations (i.e., low Knudsen numbers), this functional dependence can be made local in space through an expansion in gradients of the hydrodynamic fields. To generate it, $f$ is written as a series expansion in powers of the gradients of $n$, $\mathbf{U}$, and $T$:
\begin{equation}
f=f^{(0)}+f^{(1)}+f^{(2)}+\cdots \;, \label{4.2}
\end{equation}
where the approximation $f^{(k)}$ is of order $k$ in spatial gradients. Moreover, we assume that collisional dissipation and spatial gradients are not coupled and hence, we consider situations where the spatial gradients are sufficiently small (low Knudsen number). Moreover, in ordering the different level of approximations in the kinetic equation, one has to characterize the magnitude of the external forces relative to the gradients as well. The scaling of the forces depends on the conditions of interest. Here, as in our previous paper \cite{GTSH12}, the friction coefficient $\gamma$ will be taken to be of zeroth order in gradients since it does not induce any flux in the system. On the other hand, as for molecular gases \cite{CC70}, gravity must have a different consideration and its magnitude is at least of first order in perturbation expansion.

According to the expansion (\ref{4.2}) for the distribution function, the Enskog collision operator and time derivative are also given in the representations
\begin{equation}
\label{4.3}
J_{\text{E}}=J_{\text{E}}^{(0)}+J_{\text{E}}^{(1)}+\cdots, \quad \partial_t=\partial_t^{(0)}+
\partial_t^{(1)}+\cdots .
\end{equation}
The coefficients in the time derivative expansion are identified by a representation of the fluxes and the collisional cooling rate in the macroscopic balance equations as a similar series through their definitions as functionals of $f$.  In addition, given that collisional dissipation and gradients are uncoupled, the different approximations $f^{(k)}$ are nonlinear functions of $\alpha$, regardless of the applicability of the corresponding hydrodynamic equations truncated at that order. In this paper, we will restrict our calculations to the Navier-Stokes hydrodynamic order (first order in spatial gradients). The Burnett hydrodynamic equations (second order in spatial gradients) for a dry granular gas have been recently obtained \cite{KGS14}.

\subsection{Zeroth-order solution: local version of the HCS}

To zeroth-order, the Enskog equation (\ref{2.14}) becomes
\begin{equation}
\partial_{t}^{(0)}f^{(0)}-\frac{\gamma}{m}
\frac{\partial}{\partial{\bf V}}\cdot {\bf V} f^{(0)}=J_{\text{E}}^{(0)}[f^{(0)},f^{(0)}], \label{4.4}
\end{equation}
where $J_{\text{E}}^{(0)}[f^{(0)},f^{(0)}]\equiv \chi J_{\text{B}}[f^{(0)},f^{(0)}]$. Note that in Eq.\ (\ref{4.4}) all spatial gradients are neglected at this lowest order. Moreover, as discussed before, upon writing Eq.\ (\ref{4.4}) it has been assumed that $g$ is taken to be of first-order in spatial gradients. The macroscopic balance equations at this order give $\partial_t^{(0)}n=\partial_t^{(0)}U_i=0$ and
\begin{equation}
\label{4.5}
\partial_t^{(0)}T=-\frac{2T}{m}\gamma -\zeta^{(0)}T,
\end{equation}
where $\zeta^{(0)}$ is the cooling rate to zeroth order. A good estimate of $\zeta^{(0)}$ is given by Eq.\ \eqref{3.21}. Since $f^{(0)}$ qualifies as a {\em normal} solution, then
\begin{eqnarray}
\label{4.7}
\partial_t^{(0)}f^{(0)}&=&\frac{\partial f^{(0)}}{\partial n}\partial_t^{(0)}n+
\frac{\partial f^{(0)}}{\partial U_i}\partial_t^{(0)}U_i+\frac{\partial
f^{(0)}}{\partial T}\partial_t^{(0)}T\nonumber\\
&=& -\left(
\frac{2\gamma}{m} +\zeta^{(0)}\right)T \frac{\partial f^{(0)}}{\partial T},
\end{eqnarray}
where in the last step we have taken into account that $f^{(0)}$ depends on ${\bf U}$
through its dependence on ${\bf V}$. Substitution of Eq.\ (\ref{4.7}) into Eq.\ (\ref{4.4}) yields
\begin{equation}
\label{4.8}
\frac{1}{2}\zeta^{(0)}
\frac{\partial}{\partial {\bf
V}}\cdot {\bf V} f^{(0)}=J_{\text{E}}^{(0)}[f^{(0)},f^{(0)}].
\end{equation}
Upon deriving Eq.\ \eqref{4.8} use has been made of the relation \eqref{3.11}. A solution to Eq.\ \eqref{4.8} is given by the local version of the time-dependent distribution function \eqref{3.9}.

\section{First order solution. Navier-Stokes transport coefficients}
\label{sec5}

The analysis to first order in the Chapman-Enskog expansion is quite similar to the one worked out in Ref.\ \cite{GTSH12}. We only display in this section the final results for the fluxes and the collisional cooling rate, with some details being given in the appendix \ref{appA}. To first order, the expressions of the pressure tensor $P_{ij}^{(1)}$ and the heat flux $\mathbf{q}^{(1)}$ are given by Eqs.\ \eqref{5.1} and \eqref{5.2}, respectively, where the transport coefficients can be expressed in the forms \eqref{5.3}. The (scaled) transport coefficients $\eta^*$, $\lambda^*$, $\kappa^*$ and $\mu^*$ are nonlinear functions of the solid volume fraction $\phi$, the (dimensionless) friction coefficient $\gamma^*$ and the coefficient of restitution $\al$. They are given by \cite{GTSH12}
\begin{equation}
\label{5.5}
\eta^*=\eta_k^*\left[1+\frac{2^{d-1}}{d+2}\phi \chi \left(1+\alpha\right)\right]+\frac{d}{d+2}\lambda^*,
\end{equation}
\begin{equation}
\label{5.6}
\lambda^*=\frac{2^{2d+1}}{\pi(d+2)}\phi^2 \chi (1+\alpha)\left(1-\frac{a_2}{16} \right),
\end{equation}
\beqa
\label{5.7}
\kappa^*&=&\kappa_{k}^*\left[1+3\frac{2^{d-2}}{d+2}\phi \chi (1+\alpha)\right]+\frac{2^{2d+1}(d-1)}{(d+2)^2\pi}
\phi^2 \chi\nonumber\\
& & \times
 (1+\alpha)\left(1+\frac{7}{16} a_2 \right),
\eeqa
\begin{equation}
\label{5.8}
\mu^*=\mu_k^*\left[1+3\frac{2^{d-2}}{d+2}\phi \chi (1+\alpha)\right].
\end{equation}
According to Eqs.\ \eqref{5.5}--\eqref{5.8}, the collision contributions to the Navier-Stokes transport coefficients do not explicitly depend on the friction coefficient $\gamma^*$ (defined by the first identity in Eq.\ \eqref{2.13}) and hence, their forms are the same as those obtained for a dry granular fluid \cite{GD99,L05}. On the other hand, as we will show below, the kinetic contributions $\eta_k^*$ and $\mu_k^*$ (which are given in terms of the solutions of first-order nonlinear differential equations) present in general a complex dependence on $\gamma^*$ while the (hydrodynamic) expression of $\kappa_k^*$ is the same as the one found for dry granular fluids \cite{GD99,L05}. The results obtained for $\eta_k^*$ and $\mu_k^*$ contrast with the ones derived in Ref.\  \cite{GTSH12} where \vicente{the dependence of $\gamma^*$ on the granular temperature $T$ was neglected and hence, the above kinetic contributions obey simple algebraic equations.} Let us consider each kinetic contribution separately.

\subsection{Kinetic contribution $\eta_k^*$}

The kinetic coefficient $\eta_k^*$ obeys the first-order differential equation
\beqa
\label{5.9}
& & -\frac{1}{2}\left(2\gamma^*+\zeta_0^*\right)\left(\eta_k^*-\gamma^*\frac{\partial \eta_k^*}{\partial \gamma^*}\right)
+\left(2\gamma^*+\nu_\eta^*\right)\eta_k^*\nonumber\\
& & =1-\frac{2^{d-2}}{d+2}(1+\alpha)
(1-3 \alpha)\phi \chi,
\eeqa
where $\zeta_0^*\equiv \zeta^{(0)}/\nu$ and
\begin{equation}
\label{5.10}
\nu_\eta^*=\frac{3}{4d}\chi \left(1-\alpha+\frac{2}{3}d\right)(1+\alpha)
\left(1+\frac{7}{16}a_2\right).
\end{equation}
Note that the term $\gamma^*\partial_\gamma^* \eta_k^*$ in Eq.\ \eqref{5.9} comes directly from the temperature dependence of $\gamma^*$ since
\beq
\label{5.10.1}
T\frac{\partial \eta_k^*}{\partial T}=-\frac{1}{2}\gamma^*\frac{\partial \eta_k^*}{\partial \gamma^*}.
\eeq
The differential equation \eqref{5.9} becomes a simple linear algebraic equation when one neglects the term $\gamma^*\partial_{\gamma^*} \eta_k^*$. In this case, the form of $\eta_k^*$ is
\beq
\label{5.11}
\eta_{k,\text{approx}}^*=\frac{1-\frac{2^{d-2}}{d+2}(1+\alpha)(1-3 \alpha)\phi \chi}{\nu_\eta^*-\frac{1}{2}(\zeta_0^*-2\gamma^*)}.
\eeq
The approximated expression \eqref{5.11} for $\eta_k^*$ was already derived in Ref.\  \cite{GTSH12}. When the term $\gamma^*\partial_{\gamma^*} \eta_k^*$ is not neglected, the general solution to Eq.\ \eqref{5.9} can be written as
\beq
\label{5.12}
\eta_k^*=C \eta_{k,0}^*+\eta_{k,\text{hyd}}^*,
\eeq
where $C$ is a constant to be determined from the initial conditions,
\beq
\label{5.13}
\eta_{k,0}^*=\exp\left[
\frac{2}{\zeta_0^*}\left(\nu_\eta^*\ln \frac{2\gamma^*+\zeta_0^*}{2\gamma^*}+\frac{\zeta_0^*}{2}
\ln \frac{2\gamma^*}{(2\gamma^*+\zeta_0^*)^2}\right)\right],
\eeq
and
\beqa
\label{5.14}
\eta_{k,\text{hyd}}^*&=&\frac{2\zeta_0^*\left(1-\frac{2^{d-2}}{d+2}(1+\alpha)(1-3 \alpha)\phi \chi\right)}{\nu_\eta^*(2\gamma^*+\zeta_0^*)^2
(2\nu_\eta^*-\zeta_0^*)}\nonumber\\
&\times & \left[\nu_\eta^*(2\gamma^*+\zeta_0^*)
+\gamma^*\left(1+\frac{2\gamma^*}{\zeta_0^*}\right)^{2\nu_\eta^*/\zeta_0^*}
\right.\nonumber\\
& \times & \left. \left(2\nu_\eta^*-\zeta_0^*\right)
{_2F_1}\left(\frac{2\nu_\eta^*}{\zeta_0^*}, \frac{2\nu_\eta^*}{\zeta_0^*},1+\frac{2\nu_\eta^*}{\zeta_0^*},-\frac{2\gamma^*}{\zeta_0^*}\right)\right],\nonumber\\
\eeqa
were $_2F_1\left(a,b;c;z\right)$ is the hypergeometric function \cite{AS72}. When $\gamma^*=0$, Eq.\ \eqref{5.14} for $\eta_{k,\text{hyd}}^*$ is consistent with the expression of the kinetic shear viscosity of a dry granular gas \cite{GD99,L05}.

\begin{figure}
{\includegraphics[width=0.85\columnwidth]{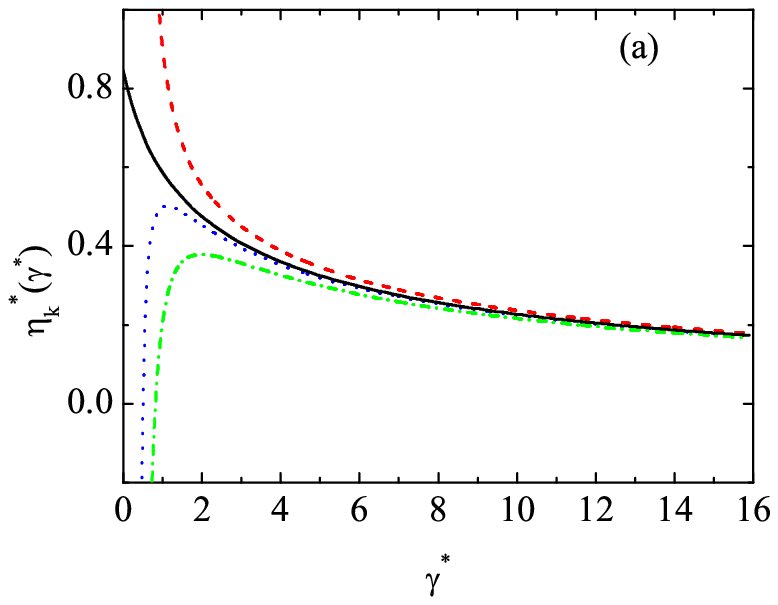}}
{\includegraphics[width=0.85\columnwidth]{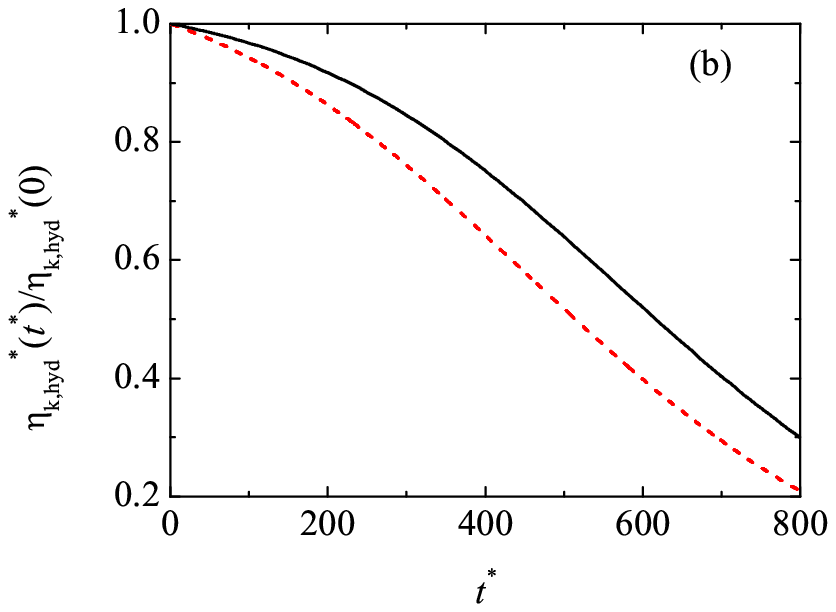}}
\caption{(Color online) (a) Plot of the kinetic contribution $\eta_k^*$ to the (dimensionless) shear viscosity as a function of the (reduced) drag coefficient $\gamma^*$ for $d=3$, $\phi=0.2$ and $\al=0.8$. Three different initial conditions are considered: $\gamma_0^*=1$ and $\eta_k^*(\gamma_0^*)=0.9$ (red dashed line), $\gamma_0^*=1$ and $\eta_k^*(\gamma_0^*)=0.5$ (blue dotted line), and $\gamma_0^*=5$ and $\eta_k^*(\gamma_0^*)=0.3$ (green dashed-dotted line). The (black) solid line corresponds to the dimensionless function $\eta_{k,\text{hyd}}^*$ (hydrodynamic shear viscosity). (b) Plot of $\eta_{k,\text{hyd}}^*(t^*)/\eta_{k,\text{hyd}}^*(0)$ versus the (dimensionless) time $t^*$ for $d=3$, $\phi=0.2$, $Re_{T_0}=5$ and $\rho_s/\rho_g=1000$. The solid and dashed lines are for $\alpha=0.9$ and $\alpha=0.8$, respectively.
\label{etainitial}}
\end{figure}

A hydrodynamic expression (independent of the initial conditions) for the shear viscosity $\eta_{k,\text{hyd}}^*$ is expected to hold after a transient period. To analyze whether the system reaches a hydrodynamic regime where $\eta_k^*=\eta_{k,\text{hyd}}^*$ we have to see if, for given values of $\alpha$, $\phi$ and $\gamma_0^*$, the ratio $\eta_{k,0}^*/\eta_{k,\text{hyd}}^*$ goes to zero for long times (which means $\gamma^*\propto (T/T_0)^{-1/2}\to \infty$ because $\nu(T)\to 0$ when $t\to \infty$ since $\nu\propto \sqrt{T}$). \vicente{Although not illustrated here, our results show that the ratio $\eta_{k,0}^*/\eta_{k,\text{hyd}}^*\to 0$ for sufficiently long times and hence, one can neglect the initial term in Eq.\ \eqref{5.12}. Moreover, the functions $\eta_{k,0}^*$ and $\eta_{k,\text{hyd}}^*$ depend both on (dimensionless) time $t^*$} through their dependence on
\beq
\label{5.15}
\gamma^*(t^*)=\frac{\gamma}{m\nu(t)}=\frac{\gamma_0^*}{\sqrt{T(t^*)/T_0}},
\eeq
where $T(t^*)/T_0$ is given by Eq.\ \eqref{3.5}. The (dimensionless) coefficient $\gamma_0^*$ can be expressed in terms of typical dimensionless numbers of suspensions, such as the ratio of the material densities of the solid and the gas phases $\rho_s/\rho_g$ and the Reynolds number $Re_{T_0}$ based on the initial granular temperature $T_0$:
\beq
\label{5.16}
Re_{T_0}=\frac{\sigma \rho_g}{\mu_g}\sqrt{\frac{T_0}{m}}.
\eeq
Note that here $Re_{T_0}$ is defined in terms of the initial temperature $T_0$ and not in terms of the time-dependent temperature $T(t)$ as in our previous work \cite{GTSH12}. The expression of $\gamma_0^*$ as a function of the Reynolds number $Re_{T_0}$ can easily be obtained when one takes into account Eqs.\ \eqref{2.5} and \eqref{5.16}. In the case of hard spheres ($d=3$), the result is
\beq
\label{5.17}
\gamma_0^*=\frac{15}{16}\frac{\sqrt{\pi}}{\phi}\frac{\rho_g}{\rho_s}\frac{R_\text{diss}(\phi)}{Re_{T_0}},
\eeq
where $\rho_s=6m/\pi \sigma^3$ for spheres. The (dimensionless) viscous dissipation function $R_\text{diss}$ was evaluated by Sangani \emph{et al.} \cite{SMTK96} as
\begin{eqnarray}
\label{5.19}
R_{\text{diss}}(\phi)&=&1+3\sqrt\frac{\phi}{2}+\frac{135}{64}\phi \ln
\phi\nonumber\\
& & +11.26 \phi \left(1-5.1 \phi+16.57 \phi^2-21.77 \phi^3\right)\nonumber\\
& & -\phi \chi(\phi)\ln \epsilon_m.
\end{eqnarray}
Equation \eqref{5.19} approaches the expression given previously by Koch \cite{K90} in the dilute limit. In Eq.\ (\ref{5.19}), $\epsilon_m \sigma$ can be interpreted as a length scale characterizing the importance of non-continuum effects on the lubrication force between two smooth particles at close contact. Typical values of the factor $\epsilon_m$ are in the range 0.01--0.05. However, since the term $\epsilon_m$ only contributes to $R_{\text{diss}}(\phi)$ through a weak logarithmic factor, its explicit value does not play a significant role in the final results. Here, we take the typical value $\epsilon_m=0.01$.

\vicente{Given that the the time dependence of the shear viscosity $\eta_k^*$ is encoded through its dependence on the (reduced) friction coefficient $\gamma^*$, to illustrate that $\eta_k^*$ achieves a hydrodynamic form, the panel (a) of Fig.\ \ref{etainitial} shows $\eta_k^*$ versus $\gamma^*$ for fixed values of $\phi$ and $\al$ and three different initial conditions (namely, different values of $\gamma_0^*$ and $\eta_k^*(\gamma_0^*)$). It is clearly seen that all the curves converge towards the universal curve $\eta_{k,\text{hyd}}^*$, which is identified as the hydrodynamic expression of the shear viscosity $\eta_k^*$. Similar conclusions have been recently found \cite{KG15} for inelastic Maxwell models of gas-solid flows. As a complement of the above plot, panel (b) of Fig.\ \ref{etainitial} shows the time dependence of the hydrodynamic form $\eta_{k,\text{hyd}}^*$ for $\phi=0.2$, $Re_{T_0}=5$ and $\rho_s/\rho_g=1000$. Two different values of the coefficient of restitution have been considered. The values of the (scaled) friction coefficient $\gamma^*$ at $t^*=800$ (the longest time considered in the plot) are $\gamma^* \simeq 8.5$ and $\gamma^* \simeq 15.5$ for $\alpha=0.9$ and 0.8, respectively. Thus, the same time scales are considered in both panels of Fig.\ \ref{etainitial}. It is quite apparent that the kinetic contribution $\eta_{k,\text{hyd}}^*$ decreases in time, being more noticeable as the collision dissipation increases}.

\begin{figure}
{\includegraphics[width=0.85\columnwidth]{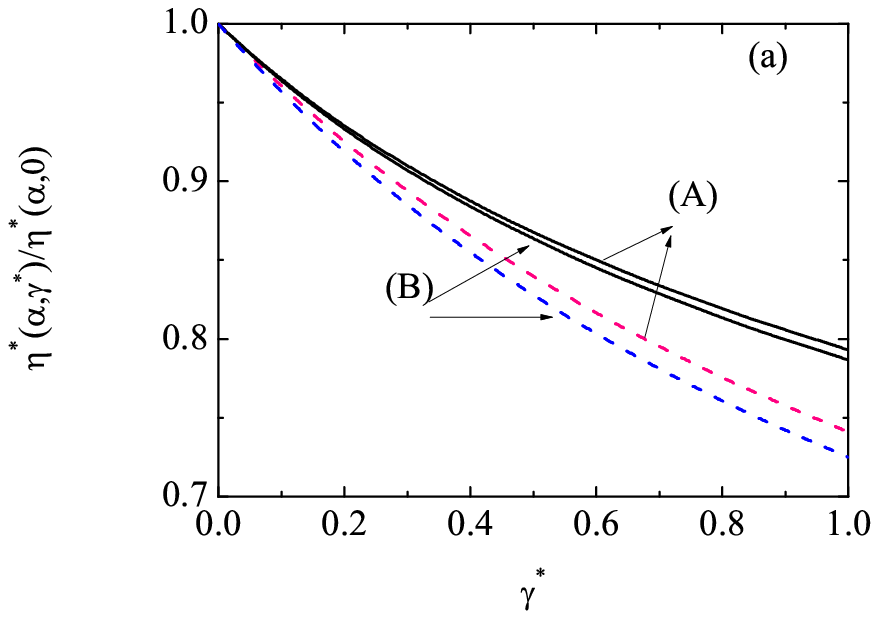}}
{\includegraphics[width=0.85\columnwidth]{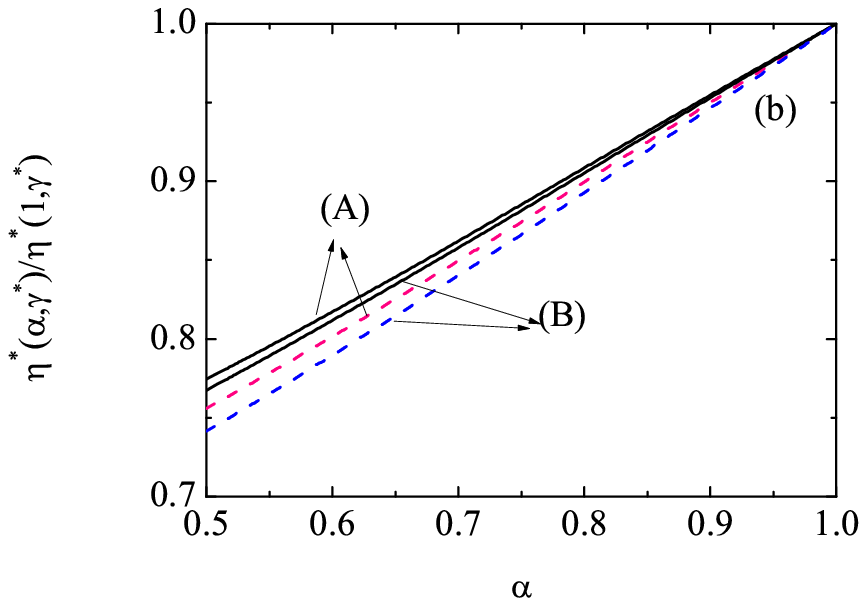}}
\caption{(Color online) (a) Plot of the ratio $\eta^*(\al,\gamma^*)/\eta^*(\alpha,0)$ versus the (dimensionless) friction coefficient $\gamma^*$ for $d=3$, $\phi=0.2$ and two different values of the coefficient of restitution $\alpha$: $\alpha=0.8$ (A) and $\alpha=0.6$ (B). (b) Plot of the ratio $\eta^*(\al,\gamma^*)/\eta^*(1,\gamma^*)$ versus the coefficient of restitution $\alpha$ for $d=3$, $\phi=0.2$ and two different values of the (dimensionless) friction coefficient $\gamma^*$: $\gamma^*=0.5$ (A) and $\gamma^*=1$ (B). In both panels, the solid lines correspond to the theoretical results derived here from Eq.\ \eqref{5.14} while the dashed lines are the (approximated) results obtained in Ref.\ \cite{GTSH12} by using Eq.\ \eqref{5.11}.
\label{eta}}
\end{figure}
\begin{figure}
{\includegraphics[width=0.85\columnwidth]{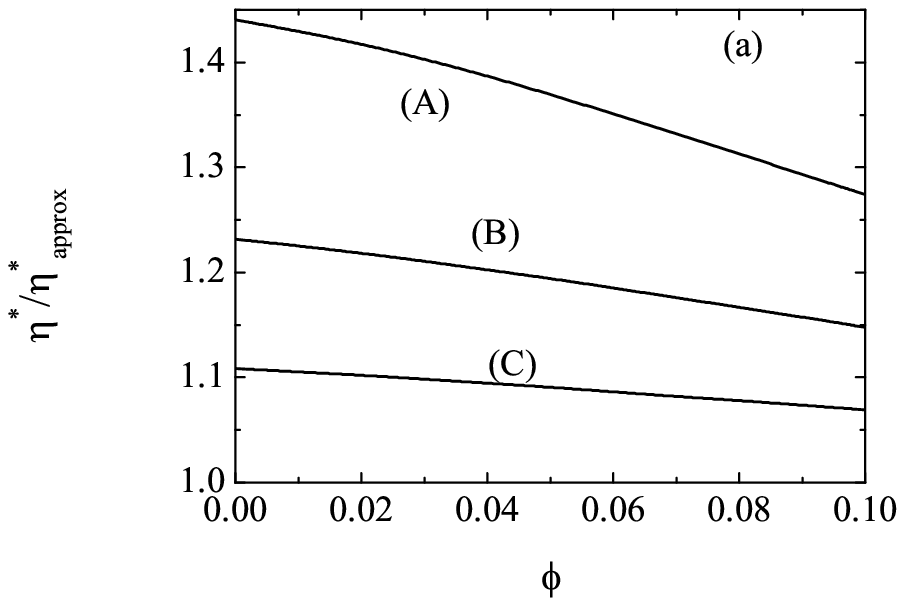}}
{\includegraphics[width=0.85\columnwidth]{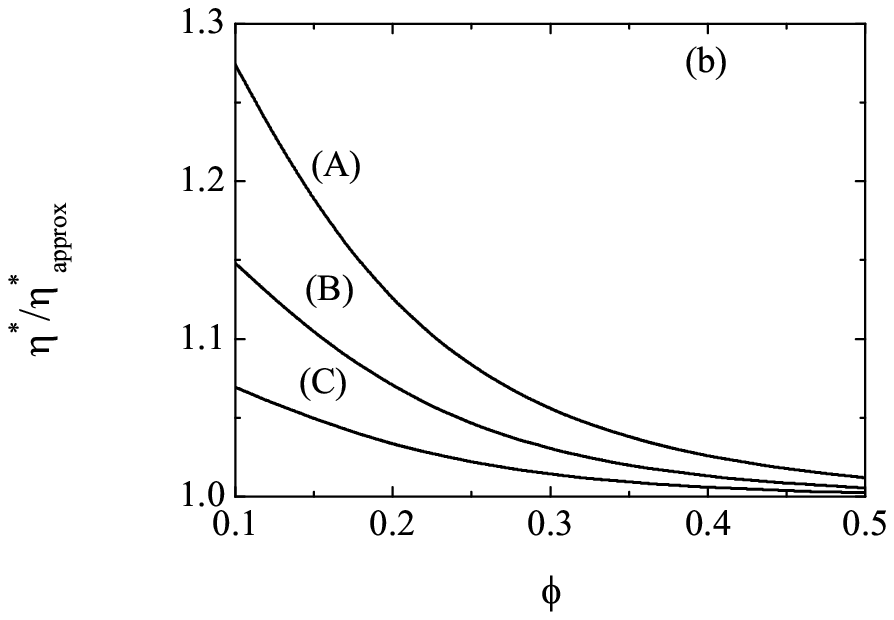}}
\caption{Plot of the ratio $\eta^*/\eta_\text{approx}^*$ versus the solid volume fraction $\phi$ for a dilute (a) and a moderately dense suspension (b) in the case of spheres ($d=3$) with $\alpha=0.8$ and three different values of the (dimensionless) friction coefficient $\gamma^*$: $\gamma^*=2$ (A), $\gamma^*=1$ (B) and $\gamma^*=0.5$ (C). In both panels, $\eta^*$ is given by Eq.\ \eqref{5.14} while $\eta_\text{approx}^*$ is the (approximated) result obtained in Ref.\ \cite{GTSH12} by using Eq.\ \eqref{5.11}.
\label{etaphi}}
\end{figure}

Once the hydrodynamic form of the kinetic contribution $\eta_{k,\text{hyd}}^*$ to the shear viscosity has been obtained, it is interesting to assess the impact of the gas phase (measured through $\gamma^*$) on the (total) shear viscosity $\eta^*$. Its expression is given by Eq.\ \eqref{5.5} with the replacement $\eta_k^* \to \eta_{k,\text{hyd}}^*$. First, the panel (a) of Fig.\ \ref{eta} shows the dependence of the ratio  $\eta^*(\al,\gamma^*)/\eta^*(1,\gamma^*)$ on $\gamma^*$ for two different values of $\alpha$. The theoretical results obtained in Ref.\ \cite{GTSH12} by using the simple form \eqref{5.11} for the kinetic contribution $\eta_k^*$ is also plotted for comparison. We observe that the predictions given by the simple expression \eqref{5.11} agree qualitatively well with those derived in this paper. On the other hand, at a more quantitative level, it is apparent that the approximated results \cite{GTSH12} for $\eta^*$ slightly underestimate the predictions found here. As a complementary plot, the panel (b) of Fig.\ \ref{eta} shows the ratio $\eta^*(\al,\gamma^*)/\eta^*(\alpha,0)$ versus $\alpha$ for different values of $\gamma^*$. As before, the agreement between the present results with those reported in our previous work \cite{GTSH12} is quite good. Finally, to assess the effect of density on the discrepancies between the results derived here and those obtained before \cite{GTSH12}, Fig.\ \ref{etaphi} shows the ratio $\eta^*/\eta_\text{approx}^*$ as a function of the volume fraction $\phi$ for $\alpha=0.8$ and three different values of $\gamma^*$. Since the kinetic contribution to $\eta^*$ (which is the only source of disagreement between both theories) dominates over its collisional contribution when $\phi \to 0$, then the discrepancies between $\eta^*$ and $\eta_\text{approx}^*$ increase as flows become more dilute. On the other hand, the above discrepancies reduce considerably as the suspension becomes denser so that the results obtained in Ref.\ \cite{GTSH12} turn out to be more accurate at moderate densities.

\subsection{Kinetic contributions $\kappa_k^*$ and $\mu_k^*$}

The kinetic contribution $\kappa_k^*$ of the (scaled) thermal conductivity coefficient verifies the differential equation
\beqa
\label{5.20}
& & (\nu_\kappa^*-2\zeta_0^*)\kappa_k^*+\frac{1}{2}\left(2\gamma^*+\zeta_0^*\right)\gamma^*\frac{\partial \kappa_k^*}{\partial \gamma^*}=\frac{d-1}{d}\left\{1+2a_2 \right. \nonumber\\
& &\left.
+3\frac{2^{d-3}}{d+2}\phi \chi(1+\alpha)^2\left[2\alpha-1+a_2(1+\alpha)\right]\right\},
\eeqa
where
\beqa
\label{5.21}
\nu_\kappa^*&=&\frac{1+\alpha}{d}\chi\left[\frac{d-1}{2}+\frac{3}{16}(d+8)(1-\alpha)\right.\nonumber\\
& & \left.+\frac{296
+217d-3(160+11d)\alpha}{256}a_2\right].
\eeqa
If one neglects the term $\partial \kappa_k^*/\partial \gamma^*$ in Eq.\ \eqref{5.20}, one gets the solution $\kappa_{k,\text{approx}}^*\equiv \kappa_{k,\text{dry}}^*$ where
\beqa
\label{5.22}
\kappa_{k,\text{dry}}^*&=&\frac{d-1}{d}\left(\nu_\kappa^*-2\zeta_0^*\right)^{-1}\left\{1+2a_2+3\frac{2^{d-3}}{d+2}\phi \chi
\right.\nonumber\\
& \times & \left.(1+\alpha)^2\left[2\alpha-1+a_2(1+\alpha)\right]\right\}
\eeqa
Note that the expression \eqref{5.22} does not depend on the friction coefficient $\gamma^*$. This means that the presence of the gas phase does not enter in the form of the thermal conductivity and hence, its expression is the same as the obtained in the dry granular case \cite{GD99,L05}. This result is consistent with one obtained in the Langevin-like model \cite{GTSH12}. The general solution to the differential equation \eqref{5.20} can be written as
\beq
\label{5.23}
\kappa_k^*=C \kappa_{k,0}^*+\kappa_{k,\text{dry}}^*,
\eeq
where $C$ is a constant to be determined from the initial conditions, $\kappa_{k,\text{dry}}^*$ is given by Eq.\ \eqref{5.22} and
\beq
\label{5.24}
\kappa_{k,0}^*=\exp \left[-\frac{2}{\zeta_0^*}\left(\nu_\kappa^*-2\zeta_0^*\right)\ln \frac{2\gamma^*}{2\gamma^*+\zeta_0^*}\right].
\eeq
As in the case of the shear viscosity, it is easy to see that, after a few collision times, the ratio $\kappa_{k,0}^*/\kappa_{k,\text{dry}}^*$ tends to zero so that, the hydrodynamic form is $\kappa_k^*=\kappa_{k,\text{dry}}^*$.

In the case of the Dufour-like coefficient, $\mu_k^*$ obeys the differential equation
\begin{widetext}
\beqa
\label{5.25}
(\nu_\kappa^*-\frac{3}{2}\zeta_0^*)\mu_k^*&+&\frac{1}{2}\left(2\gamma^*+\zeta_0^*\right)\gamma^*\frac{\partial \mu_k^*}{\partial \gamma^*}-2\kappa_k^*\gamma^*\phi \partial_\phi \ln R_\text{diss}(\phi)=
\kappa_k^* \zeta_0^{*}\left(1+\phi\partial_{\phi}
\ln \chi \right)+\frac{d-1}{d}a_2\nonumber\\
& & +3\frac{2^{d-2}(d-1)}{d(d+2)}\phi \chi
(1+\alpha)\left(1+\frac{1}{2}\phi\partial_\phi\ln\chi\right)
\left[\alpha(\alpha-1)+\frac{a_2}{6}(10+2d-3\alpha+3\alpha^2)\right].
\eeqa
As for the previous coefficients, if one neglects the term $\partial_{\gamma^*}\mu_k^*$ in Eq.\ \eqref{5.25} one gets the solution \cite{GTSH12}
\beqa
\label{5.26}
\mu_{k,\text{approx}}^*&=&\left(\nu_\kappa^*-\frac{3}{2}\zeta_0^*\right)^{-1}\left\{2\kappa_k^*\gamma^*
\phi \partial_\phi \ln R_\text{diss}(\phi)+  \kappa_k^* \zeta_0^{*}\left(1+\phi\partial_{\phi}
\ln \chi \right)+\frac{d-1}{d}a_2\right.\nonumber\\
& & \left.+ 3\frac{2^{d-2}(d-1)}{d(d+2)}\phi \chi
(1+\alpha)\left(1+\frac{1}{2}\phi\partial_\phi\ln\chi\right)
\left[\alpha(\alpha-1)+\frac{a_2}{6}(10+2d-3\alpha+3\alpha^2)\right]\right\}.
\eeqa
%
\end{widetext}
The general solution to the differential equation \eqref{5.25} is
\beq
\label{5.27}
\mu_k^*=C \mu_{k,0}^*+\mu_{k,\text{hyd}}^*,
\eeq
where $C$ is a constant,
\beq
\label{5.28}
\mu_{k,0}^*=\exp \left[-\frac{2}{\zeta_0^*}\left(\nu_\kappa^*-\frac{3}{2}\zeta_0^*\right)
\ln \frac{2\gamma^*}{2\gamma^*+\zeta_0^*}\right],
\eeq
and the explicit form of $\mu_{k,\text{hyd}}^*$ can be found in the Appendix \ref{appB}.
\begin{figure}
{\includegraphics[width=0.85\columnwidth]{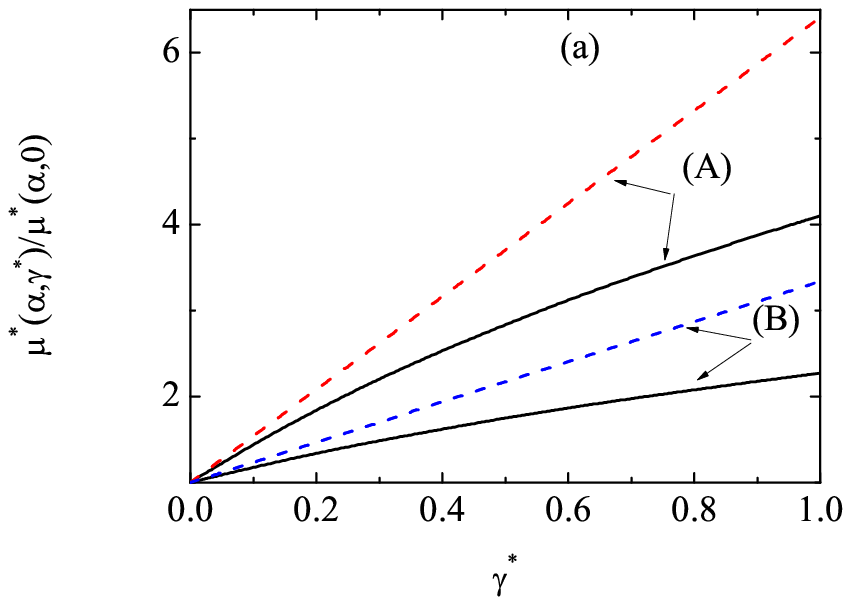}}
{\includegraphics[width=0.85\columnwidth]{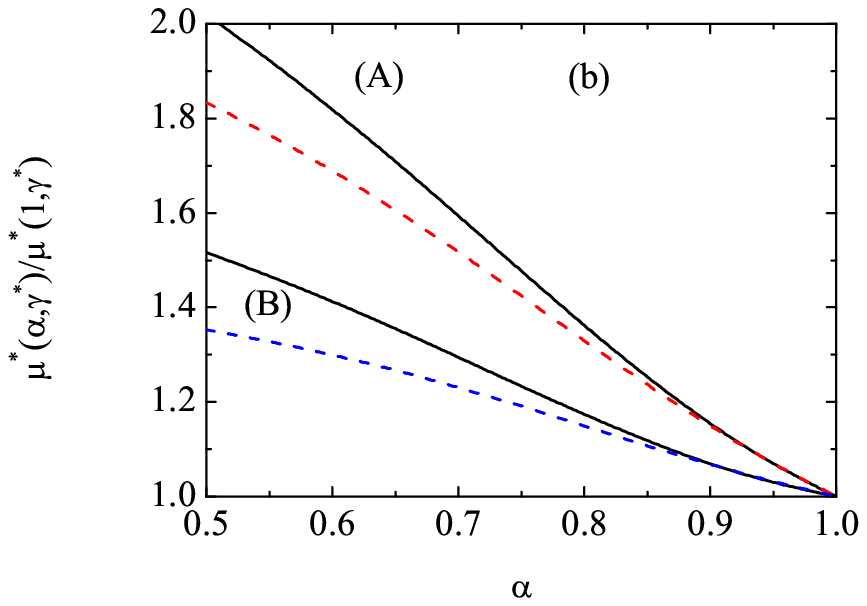}}
\caption{(Color online) (a) Plot of the ratio $\mu^*(\al,\gamma^*)/\mu^*(\alpha,0)$ versus the (dimensionless) friction coefficient $\gamma^*$ for $d=3$, $\phi=0.2$ and two different values of the coefficient of restitution $\alpha$: $\alpha=0.8$ (a) and $\alpha=0.6$ (b). (b) Plot of the ratio $\mu^*(\al,\gamma^*)/\mu^*(1,\gamma^*)$ versus the coefficient of restitution $\alpha$ for $d=3$, $\phi=0.2$ and two different values of the (dimensionless) friction coefficient $\gamma^*$: $\gamma^*=0.5$ (a) and $\gamma^*=1$ (b). In both panels, the solid lines correspond to the theoretical results derived here from \eqref{5.29} while the dashed lines are the (approximated) results obtained in Ref.\ \cite{GTSH12} by using Eq.\ \eqref{5.26}.
\label{mu}}
\end{figure}
Note that upon deriving the solution \eqref{5.27} we have assumed that the thermal conductivity has reached its hydrodynamic form $\kappa_{k,\text{dry}}^*$. As before, for long times, $\mu_{k,0}^*/\mu_{k,\text{hyd}}^*\to 0$ and hence, the hydrodynamic expression of the (dimensionless) kinetic contribution to the Dufour-like coefficient is given by Eqs.\ \eqref{5.29} and \eqref{5.30}. As in the case of $\eta_{k,\text{hyd}}^*$, the expression of $\mu_{k,\text{hyd}}^*$ agrees with the one derived for a dry granular gas when $\gamma^*=0$.

Note that if one neglects the dependence of $\gamma$ on the solid volume fraction $\phi$ in Eq.\ \eqref{2.5} (which is equivalent to assume $R_{\text{diss}}\equiv \text{const.}$ and so, the quantity $B=0$ in Eq.\ \eqref{5.31}), then the form of $\mu_{k,\text{hyd}}^*$ is the same as the one obtained for a dry granular gas \cite{GD99,L05}. This result is consistent with the mapping proposed in Ref.\ \cite{L01} when $\gamma$ is an arbitrary constant since in this simple case the expressions of $\kappa_{k,\text{hyd}}^*$ and $\mu_{k,\text{hyd}}^*$ with and without the drag force are the same. However, even in this case ($\gamma\equiv \text{const.}$) the hydrodynamic form of the (dimensionless) kinetic contribution to the shear viscosity [see Eq.\ \eqref{5.14}] still presents a complex dependence on $\gamma^*$ and hence, there is not an exact equivalence between both descriptions (with and without the external drag force). A possible reason for the discrepancy between our results for $\eta_k^*$ and those obtained in Ref.\ \cite{L01} could stem from the different form of the drag force since the latter work considered a drag force proportional to the particle velocity $\mathbf{v}$ instead of the peculiar velocity $\mathbf{V}(\mathbf{r},t)=\mathbf{v}-\mathbf{U}(\mathbf{r},t)$ considered here. In fact, previous results \cite{DSBR86,GS03} derived for ordinary (elastic) gases under uniform shear flow have shown that a drag force of the form $-\gamma \mathbf{V}$ generally does not play a neutral role in the nonlinear rheological properties of the gas (except for the special case of Maxwell molecules) and hence, these properties are different from those derived in the absence of the drag force. On the other hand, it could be perhaps possible that the explicit dependence of $\eta_k^*$ on $\gamma^*$ of Eq.\ \eqref{5.14} could be eliminated by employing the coordinates proposed in Ref.\ \cite{L01} (such as the logarithmic time scale introduced in this paper) instead of the the conventional (reduced) time scale $t^*\equiv \nu(T_0)t$. Here, we have preferred to use the original form of the dynamics in order to maintain consistency with the simulation results presented in sections \ref{sec7} and \ref{sec8}.

Another interesting limit is $\phi \to 0$ (very dilute suspensions). In this case, the collisional contribution to $\mu^*$ vanishes and hence, $\mu^* \simeq \mu_k^*=\mu_{k,\text{hyd}}^*$ where
\beq
\label{5.33}
\mu_{k,\text{hyd}}^*=\frac{\kappa^* \zeta_0^{*}+\frac{d-1}{d}a_2}{\nu_\kappa^*-\frac{3}{2}\zeta_0^*}.
\eeq
Equation \eqref{5.33} coincides with the dilute limit of the approximated form \eqref{5.26}, which is also independent of the (scaled) friction coefficient $\gamma^*$.

The panels (a) and (b) of Fig.\  \ref{mu} show a comparison between the results derived here for $\mu^*$ with those obtained by using Eq.\ \eqref{5.26}. The explicit form of $\mu^*$ is given by Eq.\ \eqref{5.8} with the change $\mu_k^*\to \mu_{k,\text{hyd}}^*$. As in the case of the shear viscosity (see Fig.\  \ref{eta}), we observe that the simple expression \eqref{5.26} for $\mu^*$ captures qualitatively well the dependence of this coefficient on both $\al$ and $\gamma^*$. However, at a more quantitative level, it is quite apparent that there are significant discrepancies between both theoretical predictions especially for high values of $\gamma^*$.
\begin{figure}
\includegraphics[width=0.85\columnwidth]{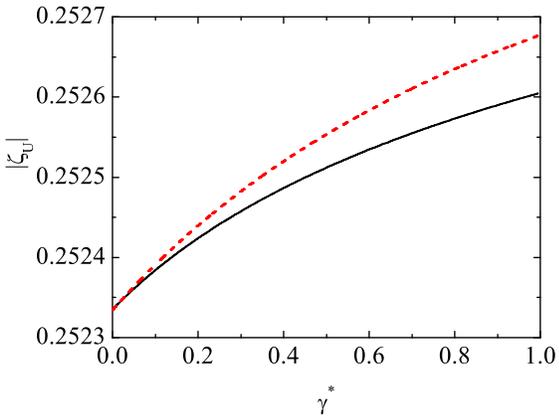}
\caption{(Color online) Plot of the magnitude of the first-order contribution $\zeta_U$ to the cooling rate versus the (dimensionless) friction coefficient $\gamma^*$ for a three-dimensional system with $\phi=0.2$ and $\al=0.8$. The solid line is the result derived here while the (red) dashed line corresponds to the expression \eqref{b12} obtained in Ref.\ \cite{GTSH12}.
\label{zetaU}}
\end{figure}

\subsection{Collisional cooling rate}

To close this section, it is important to recall that the collisional cooling rate $\zeta$ has a first-order contribution proportional to the divergence of flow velocity $\nabla \cdot \mathbf{U}$. To first order in spatial gradients, the collisional cooling rate is given by
\beq
\label{5.34}
\zeta=\zeta^{(0)}+\zeta_U \nabla \cdot \mathbf{U},
\eeq
where $\zeta^{(0)}$ is defined by Eq.\ \eqref{3.21}. The expressions of the Euler transport coefficient $\zeta_U$ is given by Eq.\ \eqref{b13}. Note that $\zeta_U$ vanishes in both for dilute suspensions ($\phi=0$) and for elastic particles ($\al=1$).
Figure \ref{zetaU} illustrates the $\gamma^*$-dependence of the magnitude $\zeta_U$. The approximated result \eqref{b13} obtained in Ref.\ \cite{GTSH12} is also plotted. It is quite apparent that the impact of the gas phase on $|\zeta_U|$ is very tiny since the magnitude of this coefficient does not change appreciable with $\gamma^*$. Moreover, the performance of the approximated expression \eqref{b12} (the form of which is much more simple than Eq.\ \eqref{b13}) is quite good in the entire range of values of $\gamma^*$ studied.

\section{Stability of the linearized hydrodynamic equations}
\label{sec6}

When the expressions of the pressure tensor, the heat flux and the cooling rate are substituted
into the balance equations \eqref{2.8}, \eqref{2.15} and \eqref{2.16} one gets the corresponding Navier-Stokes (closed) hydrodynamic equations for the number density $n$, the flow velocity ${\bf U}$ and the granular temperature $T$. In the absence of gravity ($\mathbf{g}=\mathbf{0}$), they are given by
\begin{equation}
\label{6.1}
D_tn+n\nabla \cdot {\bf U}=0,
\end{equation}
\beqa
\label{6.2}
& &D_t U_i+(nm)^{-1}\nabla_i p=(nm)^{-1}\nabla_j\left[\eta \left(\nabla_i U_j+\nabla_j U_i\right.\right.\nonumber\\
& &
\left.\left.-\frac{2}{d} \delta_{ij}\nabla \cdot {\bf U}\right)+\lambda \delta_{ij}\nabla \cdot {\bf U}\right],
\eeqa
\begin{eqnarray}
\label{6.3}
& & n\left(D_t+\frac{2\gamma}{m}+\zeta^{(0)}\right)T+\frac{2}{d}p\nabla \cdot {\bf U}=\frac{2}{d}\nabla \cdot \left
(\kappa \nabla T\right.\nonumber\\
& &\left.+\mu \nabla n\right) +\frac{2}{d}\left[\eta \left(\nabla_i U_j+\nabla_j U_i-\frac{2}{d}
\delta_{ij}\nabla \cdot {\bf U}\right)\right.\nonumber\\
& & \left.+\lambda \delta_{ij}\nabla \cdot {\bf U}\right]\nabla_i U_j
-nT\zeta_U  \nabla\cdot {\bf U}.
\end{eqnarray}
Note that consistency would require to consider up to second order in the gradients in the expression \eqref{5.34}
for the cooling rate, since this is the order of the terms in Eq.\ \eqref{6.3} coming from
the pressure tensor and the heat flux. However, it has been shown for a granular dilute gas that the contributions from the cooling rate of second order are negligible as compared with the corresponding contributions from Eqs.\ \eqref{5.1} and \eqref{5.2} \cite{BDKS98}. A similar behavior is expected in the case of suspensions at moderate densities.

As analyzed in Sec.\ \ref{sec3}, the hydrodynamic equations \eqref{6.1}--\eqref{6.3} admit a simple solution which corresponds to the so-called HCS. Nevertheless, this homogeneous time-dependent state is expected to be unstable with respect to long enough wavelength perturbations as some computer simulations have previously shown for granular \cite{GZ93,M93,MY94,MY96,BRC99b,MDCPH11,MGHEH12,MGH14} and gas-solid \cite{WK00} flows. We want here to analyze the stability of the HCS of the suspension by using the drag model \eqref{2.14}. In order to study this problem it is convenient to carry on a (linear) stability analysis of the nonlinear hydrodynamic equations \eqref{6.1}--\eqref{6.3} with respect to the homogeneous state for small initial excitations. As expected, the linearization of the Navier-Stokes hydrodynamic equations about the homogeneous solution yields partial differential equations with coefficients that are independent of space but depend on time since the reference state is cooling. However, in contrast to previous stability analysis \cite{BDKS98,G05,GMD06} for (dry) granular gases, the time dependence of the above coefficients cannot completely be eliminated after changing the time and space variables and scaling the hydrodynamic fields due essentially to the different time scale of the drag parameter $\gamma$ and the remaining time dependent parameters involved in the problem. As will show, this fact introduces additional difficulties not present in previous works \cite{BDKS98,G05,GMD06}.

Let $\delta y_{\beta}(\mathbf{r},t)=y_{\beta}(\mathbf{r},t)-y_{H\beta}(t)$ denote the deviation of $\{n, \mathbf{U}, T\}$ from their values in the HCS. In this case, the hydrodynamic fields can be written as
\begin{subequations}
\begin{equation}
\label{6.4}
n(\mathbf{r},t)=n_H+\delta n (\mathbf{r},t), \quad \mathbf{U}(\mathbf{r},t)=\delta \mathbf{U}(\mathbf{r},t),
\eeq
\beq
T(\mathbf{r},t)=T_H(t)+\delta T (\mathbf{r},t),
\end{equation}
\end{subequations}
where the quantities in the homogeneous state verify $\nabla n_H=\nabla T_H=0$ and the granular temperature $T_H$ is given by Eq.\ \eqref{3.5} (or Eq.\ \eqref{3.21.1} in terms of $\tau$). If the spatial perturbation is sufficiently small, then for some initial time interval these deviations will remain small and the hydrodynamic equations \eqref{6.1}--\eqref{6.3} can be linearized with respect to $\delta y_{\beta}(\mathbf{r},t)$. As in previous studies \cite{BDKS98,G05,GMD06}, we consider the (dimensionless) time variable $\tau$ defined by the relation \eqref{3.18} and introduce the (dimensionless) space variable
\begin{equation}
\label{6.5}
\boldsymbol{s}= \frac{1}{2}\frac{\nu_H(t)}{v_H(t)}\mathbf{r},
\end{equation}
where $\nu_H(t)$ is defined by Eq.\ \eqref{3.6} and $v_H(t)=\sqrt{T_H(t)/m}$. According to Eq.\ \eqref{6.5}, the unit length $\nu_H(t)/v_H(t)$ is proportional to the effective time-independent mean free path $1/n_H\sigma^{d-1}$.

A set of Fourier transformed dimensionless variables are then
introduced by
\begin{equation}
\label{6.6}
\rho_{{\bf k}}(\tau)=\frac{\delta n_{{\bf k}}(\tau)}{n_{H}}, \quad
{\bf w}_{{\bf k}}(\tau)=\frac{\delta {\bf U}_{{\bf k}}(\tau)}{v_H(\tau)},\quad
\theta_{{\bf k}}(\tau)=\frac{\delta T_{{\bf k}}(\tau)}{T_{H}(\tau)},
\end{equation}
where $\delta y_{{\bf k}\alpha}\equiv \{\delta n_{{\bf k}},{\bf
w}_{{\bf k}}(\tau), \theta_{{\bf k}}(\tau)\}$ is defined as
\begin{equation}
\label{6.7}
\delta y_{{\bf k}\alpha}(\tau)=\int \dd {\boldsymbol {s}}\;
e^{-i{\bf k}\cdot {\boldsymbol {s}}}\delta y_{\alpha}
({\boldsymbol {s}},\tau).
\end{equation}
Note that in Eq.\ \eqref{6.7} the wave vector ${\bf k}$ is dimensionless.

As expected, the transverse velocity components ${\bf w}_{{\bf k}\perp}={\bf w}_{{\bf k}}-({\bf w}_{{\bf k}}\cdot
\widehat{{\bf k}})\widehat{{\bf k}}$ (orthogonal to the wave vector ${\bf k}$) decouple from the other three modes and hence can be obtained more easily. Their evolution equation is
\begin{equation}
\label{6.8}
\frac{\partial {w}_{{\bf k}\perp}}{\partial \tau}-\left(2\gamma^*+\zeta_0^*-\frac{1}{2}\eta^*
k^2\right){w}_{{k}\perp}=0,
\end{equation}
where $\zeta_0^*\equiv \zeta_H^{(0)}/\nu_H$, $\gamma^*\equiv \gamma/(m \nu_H)$ and $\eta^*\equiv \eta_H/(n_H T_H/\nu_H)$. All these quantities are understood that they are evaluated in the reference base state (HCS). Note that in Eq.\ \eqref{6.8}, $\gamma^*$ and $\eta^*$ are still time dependent functions. In the granular limit case ($\gamma^*=0$), $\eta^*$ is independent of time and Eq.\ \eqref{6.8} becomes a simple (linear) differential equation whose solution is
\beq
\label{6.9}
{w}_{{\bf k}\perp,\text{dry}}(\tau)={w}_{{\bf k}\perp}(0) \exp\left[\left(\zeta_0^*-\frac{1}{2}\eta^* k^2\right)\tau\right].
\eeq
Equation \eqref{6.9} shows that there exists a critical wavenumber $k_{c,\perp}^{\text{dry}}=\sqrt{2\zeta_0^*/\eta^*}$ such that the transversal shear modes become unstable when $k<k_{c,\perp}^{\text{dry}}$. On the other hand, beyond the granular limit case, the differential equation \eqref{6.9} must numerically be integrated. Although decoupled, the transverse velocity mode is set as the fourth component of the perturbation vector and solved alongside the remaining (longitudinal) modes for numerical convenience.

The longitudinal modes correspond to $\rho_{{\bf k}}$, $\theta_{{\bf k}}$, and the longitudinal velocity component of the velocity field, $w_{{\bf k}||}={\bf w}_{{\bf k}}\cdot \widehat{{\bf k}}$ (parallel to ${\bf k}$). These modes are coupled and obey the equation
\begin{equation}
\frac{\partial \delta y_{{\bf k}\alpha }(\tau )}{\partial \tau }=M_{\alpha \beta}
 \delta y_{{\bf k}\beta }(\tau ),
\label{6.10}
\end{equation}
for the first three components of the perturbation vector which now denotes the set $\delta y_{{\bf k}\alpha }(\tau ) =
\left\{ \rho _{{\bf k}},\theta _{{\bf k}}, w_{{\bf k}||} \right\}$ and ${\sf M}$ is the square matrix
\begin{widetext}
\begin{equation}
{\sf M}=\left(
\begin{array}{ccc}
0 & 0 & -i k \\
-2\left(\zeta_0^*g+2\gamma_n^*\right)-\frac{d+2}{4}\mu^*k^2&-\zeta_0^*-\frac{d+2}{4}\kappa^*k^2 & -ik(\frac{2}{d}p^*+\zeta_U)\\
-ikp^*C_\rho & -ikp^* &\zeta_0^{*}+2\gamma^*-(\frac{d-1}{d}\eta^*+\frac{1}{2}\lambda^*)k^2
\end{array}
\right).   \label{6.11}
\end{equation}
\end{widetext}
As before, it is understood that $\eta^*$, $\lambda^*$, $\kappa^*$, $\mu^*$,
$\zeta_0^*$, and $\zeta_U$ are evaluated in the HCS. Furthermore, the quantities $p^*$, $g(\phi)$, $\gamma_n^*$ and
$C_\rho(\alpha,\phi)$ are given, respectively, by
\beq
\label{pressure}
p^*\equiv \frac{p_H}{n_H T_H}=1+2^{d-2}(1+\alpha)\chi \phi,
\eeq
\begin{equation}
\label{6.12}
g(\phi)=1+\phi\frac{\partial}{\partial \phi}\ln \chi(\phi),
\end{equation}
\beq
\label{6.13}
\gamma_n^*(\phi)=\gamma^* \phi \frac{\partial}{\partial \phi}\ln R_\text{diss}(\phi),
\eeq
\beq
\label{6.14}
C_\rho(\alpha,\phi)=1+g(\phi)\frac{p^*(\alpha,\phi)-1}{p^*(\alpha,\phi)}.
\eeq
In the absence of the gas phase ($\gamma^*=0$), the matrix equation \eqref{6.10} is consistent with previous results derived for granular gases \cite{G05}. In addition, in this limit case, the longitudinal modes become unstable for $k<k_{c,||}^{\text{dry}}$ where
\beq
\label{6.14.1}
k_{c,||}^{\text{dry}}=\sqrt{\frac{4}{d+2}\frac{2g-C_\rho}{C_\rho \kappa^*-\mu^*}}.
\eeq
As in the case of the transversal shear modes, the time dependence of the longitudinal modes must be obtained by numerically integrating Eq.\ \eqref{6.10}. The standard four--step, fourth--order Runge--Kutta method is used for numerical integration with four modes solved together.

\begin{figure}
\centering
\includegraphics[width=0.85\columnwidth]{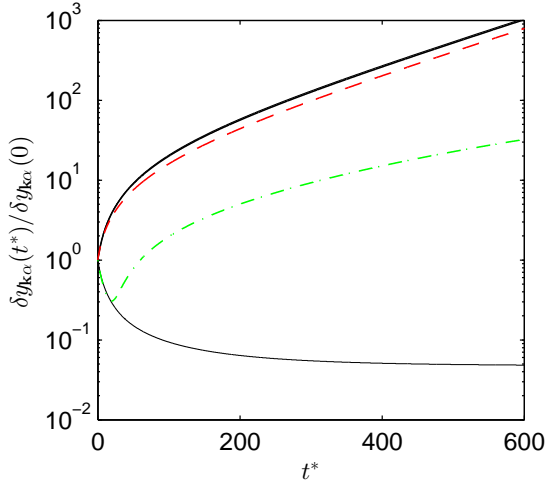}
\caption{(Color online) Dependence of the (dimensionless) perturbation variables on the (dimensionless) time $t^*$ for a large ($L^* = 100$) three-dimensional system ($d=3$) with $\phi=0.2$, $Re_{T_0}=5$, $\rho_s/\rho_g=1000$ and $\alpha=0.8$. The numerical solutions for the transverse velocity (black solid line), the longitudinal velocity (red dashed line), and the temperature (green dash-dotted line) modes are compared to the analytical Euler limit (thin lines), i.e., $L^* \to \infty$, see Eqs.\ \eqref{6.14.2} and \eqref{6.14.3}. Note that the transverse velocity mode has collapsed to its Euler limit and is indistinguishable in the plot.
\label{lsa0}}
\end{figure}

Before analyzing the general case, it is instructive to consider first the solutions to Eqs.\ \eqref{6.8} and \eqref{6.10} in the extreme long wavelength limit (i.e., $k=0$). It can be seen from Eq.\ \eqref{6.8} that the long wavelength limit gives the most unstable solution. This situation corresponds to evolution of the suspension due to uniform perturbations of the HCS, i.e., a global change in the HCS parameters. In this limit case (Euler hydrodynamics), the density $\rho _{{\bf k}}(\tau)\equiv \text{const.}$, the transverse and longitudinal velocity modes are degenerate ($w_{{\bf k}\perp}(\tau)=w_{{\bf k}||}(\tau)\equiv w_{{\bf k}}(\tau)$) and given by
\beq
\label{6.14.2}
w_{{\bf k}}(\tau)=w_{{\bf k}}(0)\frac{\zeta_0^* e^{\zeta_0^* \tau}}{2\left(1-e^{\zeta_0^* \tau}\right)\gamma_0^*+\zeta_0^*},
\eeq
while the temperature $\theta _{{\bf k}}(\tau)$ is
\beq
\label{6.14.3}
\theta_{{\bf k}}(\tau)=\theta_{{\bf k}}(0)e^{-\zeta_0^* \tau}.
\eeq
Note that upon deriving Eq.\ \eqref{6.14.3} we have taken the initial condition $\rho _{{\bf k}}(0)=0$ for the sake of simplicity. Equation \eqref{6.14.3} clearly shows that the temperature is a decaying mode and hence, it is stable. On the other hand, an analysis of the time dependence of the shear modes shows that both modes are unstable. The analytical Euler expressions are compared with the numerically integrated solutions of a large system in Fig.\ \ref{lsa0} for hard spheres ($d=3$) with $\phi=0.2$, $Re_{T_0}=5$, $\rho_s/\rho_g=1000$ and $\alpha=0.8$. In Fig.\ \ref{lsa0}, the numerically integrated solution for transverse mode is almost completely coincident with \vicente{its} Euler limit form \eqref{6.14.2}, while the longitudinal mode grows at a slightly lower rate. The temperature mode begins to follow the form \eqref{6.14.3}, but beyond $t^* \simeq 20$ the $k$-dependent numerical solution begins to diverge. The time period for which the numerical solution agrees with the analytical solution increases with increasing the (scaled) system size, \vicente{$L^*\equiv L/\sigma$}.

In a system of finite size with periodic boundary conditions, the smallest allowable wavenumber is $2\pi/L$, where $L$ is the largest system length, or, in dimensionless units,
\beq
\label{6.15}
k_\text{min} = \frac{2 v_H(t)}{\nu_H(t)} \frac{2\pi}{L} = \frac{\pi^{3/2}(d+2)}{2^d\;d}\left(\phi L^*\right)^{-1}.
\eeq
Since the smallest allowable wavenumber is the most unstable wavenumber, $k_\text{min}$ is the only wavenumber considered in \vicente{our} analysis.
\begin{figure}
\centering
\includegraphics[width=0.85\columnwidth]{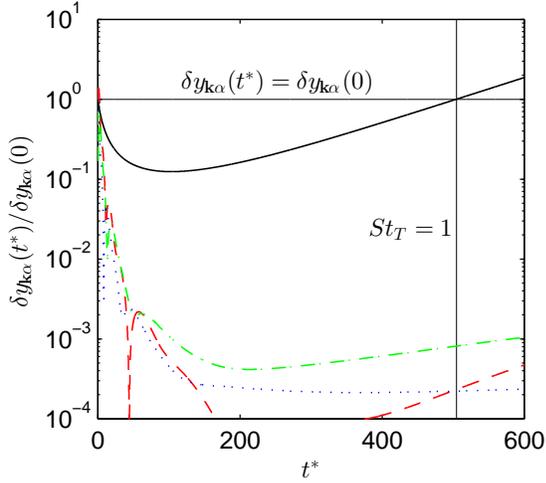}
\caption{(Color online) Dependence of the (dimensionless) perturbation variables on the (dimensionless) time $t^*$ for a critical ($L^* = L^*_\text{crit}$) three-dimensional system ($d=3$) with $\phi=0.2$, $Re_{T_0}=5$, $\rho_s/\rho_g=1000$ and $\alpha=0.8$. The numerical solutions for the transverse velocity (black solid line), the longitudinal velocity  (red dashed line), the temperature (green dash-dotted line) and the density (blue dotted lines) modes are compared.
\label{lsa1}}
\end{figure}

Opposite the Euler limit, in the case of vanishing domain size ($k \to \infty$) the system becomes stable as clearly shown in Eq.\ \eqref{6.8}. The goal then is to determine the wavenumber (system size) at which critical or neutral stability is achieved. However, unlike in the granular system, \vicente{the perturbation variables are time-dependent functions and hence}, stability cannot be simply determined from the behavior at $\tau = 0$. As shown in Eq.\ \eqref{5.15}, $\gamma^*$ is inversely proportional to the granular temperature so that $\gamma^*$ increases \vicente{with} time. In terms of linear stability, increasing $\gamma^*$ causes systems that are stable at $\tau = 0$ to be unstable at a later time.

The time dependence of \vicente{the (scaled) friction coefficient} $\gamma^*$ also has the unfortunate consequence of causing even unrealistically small systems (i.e., $L < \sigma$) to becomes unstable at some point in time. \vicente{In addition, the solid shear viscosity} also depends on time through its dependence on the \vicente{$\gamma^*$}, however this effect is considerably smaller than the changes to $\gamma^*$ directly and only exacerbates the problem. This challenge is overcome by noting that the \vicente{linear stability analysis is expected to be reliable only in the first stages of the evolution of the system}. While lubrication forces were considered in the DNS simulations conducted in deriving the thermal drag model, at a specified particle separation distance the lubrication force model is truncated to avoid the singular limit at contact. The effect of this separation distance appears in the resulting thermal drag model as $\epsilon_m$, see Eq.\ \eqref{5.19}. To understand what happens very close to particle contact, one must consider non-continuum effects. Using the \wdf{linearized Boltzmann equation for incompressible molecular flows}, Sundararajakumar and Koch \cite{SK96} have determined a critical Stokes number below which colliding particles will not have enough inertia to overcome the lubrication force and return to their original positions. \wdf{In other words, there is another condition external to the present stability analysis that will effectively end the HCS at a finite $t^*$.} By taking the relative velocity of colliding particles to be twice the thermal speed, i.e., $2\sqrt{2T/m}$, the critical Stokes number is given by
\beq
\label{6.16.1}
St_{T,\text{crit}} = \frac{1}{2\sqrt{2}} \left[ \ln \left( \frac{\epsilon_m \sigma}{2 \ell_g} \right) - 1.28 \right],
\eeq
where $\ell_g$ is the mean free path of the interstitial molecular gas and
\beq
\label{6.16.2}
St_T \equiv \frac{2 \sqrt{m T}}{3 \pi \mu_g \sigma^2} = \frac{1}{9} \frac{\rho_s}{\rho_g} Re_T
\eeq
is the thermal Stokes number. \vicente{In Eq.\ \eqref{6.16.2}, $Re_T$ is given by Eq.\ \eqref{5.16} by replacing $T_0 \to T$.}

Considering conditions relevant to circulating fluidized bed applications, we take the fluidizing medium to be air, $\ell_g = 68$~nm, and a range of practical grain sizes to be $\sigma =$~0.1--1~mm. Then, taking $\epsilon_m = 0.01$ as before, Eq.\ \eqref{6.16.1} gives a range of critical Stokes numbers of 0.25--1.07. The value of the \vicente{(scaled) friction coefficient $\gamma^*$} at $St_T = St_{T,\text{crit}}$ can be determined by rearranging Eq.\ \eqref{6.16.2} and inserting into Eq.\ \eqref{5.15}. Taking the more conservative (i.e. breakdown occurs sooner) value of $St_{T,\text{crit}} = 1$, the critical value  \vicente{$\gamma^*_\text{crit}$ of the friction coefficient} becomes
\beq
\label{6.16.3}
\gamma^*_\text{crit} = \frac{5 \sqrt{\pi}}{48} \frac{R_{\text{diss}}(\phi)}{\phi} .
\eeq
Since $\gamma^*_\text{crit}$ depends only on the concentration, it is also possible to determine its minimum value, 3.6, which occurs at approximately $\phi = 0.23$. Even at this minimum critical value, this appears to be a sufficiently long time for the collapse of the time dependent shear viscosity into \vicente{its} hydrodynamic solution, as evidenced by \vicente{the panel (a) of Fig.\ \ref{etainitial}}. \wdf{Therefore, the stability of a given system will be classified based on its behavior at this critical time, i.e., when $\gamma^* = \gamma^*_\text{crit}$.}

In contrast to the \vicente{\emph{dry}} granular case where the stability of a system was classified at the initial state, now stability is determined at the final state. While there are now several options for stability criteria, we choose a simple one: that the magnitude of a perturbation mode be larger than its initial value \wdf{at the critical time}. In other words, a system is said to be unstable if $\delta y_{{\bf k}\alpha}(\tau) \ge \delta y_{{\bf k}\alpha}(0)$ when $St_T(\tau) \ge St_{T,\text{crit}}$ \wdf{(or equivalently when $\gamma^* \le \gamma^*_\text{crit}$)} and neutrally stable at the coincidence of their equalities. One such neutrally stable system is shown in Fig.\ \ref{lsa1} for the same conditions as in Fig.\ \ref{lsa0}, for which $L^*_\text{crit} \approx 7.4$. For this case and in the following analysis, the initial condition of each mode is set to unity. As in the dry granular case \cite{G05,MDCPH11}, the transverse velocity component is the most unstable mode. While the longitudinal velocity mode is slightly larger than its initial value for a very short time, it quickly becomes stable and decays with the remaining longitudinal modes. Similar results were obtained throughout the parameter space. Therefore, only the transverse velocity mode is considered for the onset of instability \wdf{in the following analysis. Qualitatively, the transverse velocity mode is connected with the velocity vortex instability (i.e. coherent particle motion) and not with the clustering instability (inhomogeneous particle distribution).}

\section{Direct Numerical Simulations}
\label{sec7}

\wdf{Previously, molecular dynamics (MD) simulations have been used to validate the theoretical predictions of the linear stability analysis of the hydrodynamic equations in \emph{dry} granular systems \cite{MDCPH11,MGHEH12,MGH14}.  In particular, the theoretical predictions of the velocity vortex instability showed excellent agreement with the numerical data \cite{MDCPH11}.  In a similar vein, a series of numerical simulations have also been performed to quantify the accuracy of the stability analysis performed in Sec.\ \ref{sec6}. The present case with an interstitial fluid presents an additional challenge. Two common numerical strategies to model both the fluid and particle dynamics are computational fluid dynamics with discrete element modeling (CFD-DEM) and DNS. While CFD-DEM is significantly more affordable, whether current drag modeling approaches accurately capture all contributions of the instantaneous fluid force remains an open issue \cite{GTSH12}. Therefore, DNS -- which does not require modeling -- is preferred, although such an approach is considerably more computationally expensive. DNS has been used previously to study instabilities in the HCS, showing good agreement with the theoretical granular temperature decay in the early stages for elastic \cite{WK00} and inelastic particles \cite{YZMH13}. However both studies focused on relatively large systems, all of which led to the development of vortex and/or clustering instabilities. Here, we perform simulations for a range of system sizes to determine the critical system size, $L^*_{\text{crit}}$, for a given set of conditions.}

\subsection{Numerical method}
The SUSP3D program, developed by Ladd and coworkers for particulate flows \cite{L94a,L94b,LV01} is used here to generate the DNS data for comparison with the theoretical results obtained from the linear stability analysis. This program uses a \wdf{three-dimensional, 19-velocity quadrature (D3Q19)} lattice Boltzmann method (LBM) to solve the fluid flow. Solid particles, represented by collections of discrete solid nodes on the cubic lattice, are moved by fluid-particle forces using the Newton's law of motion. Collisions between particles are treated as hard-sphere collisions with a normal coefficient of restitution. Lubrication interactions between particles are supplied by analytical models and the singularity at contact is resolved by applying the lubrication cut-off \cite{NL02}.

\subsection{Setup of simulations}
DNS of the HCS used cubic periodic computational domains. The size of the computational domain, $L$, is determined by the dimensionless domain size of interest $L^* \equiv L/\sigma$, and the minimum lattice resolution needed to resolve flows around particles. In this study, $\sigma$ is taken to be 5.84$\Delta x$, where $\Delta x$ is the lattice spacing. This lattice resolution was tested against a higher resolution 9.84$\Delta x$  and the results were identical, indicating that \wdf{a resolution of $\sigma = 5.84 \Delta x$} is sufficient. The number of particles in the HCS is determined by the solid fraction; initially the particles are distributed randomly in the computational domain.

All simulations started from an initial condition in which the fluid is stationary and the particle velocities follow a Gaussian velocity distribution with a zero mean. The standard deviation of the velocity distribution gives $Re_{T_0}$. \wdf{As the particle velocities are determined randomly, $Re_{T_0}$ of each replicate is not identical and may vary from the reported values by $\pm$10\%}. The system is characterized by the combination of five nondimensional variables: $L^*$, $\phi$, $\alpha$, $Re_{T_0}$, and $\rho_s/\rho_g$. \wdf{Five simulations are carried out for each condition studied, each replicate with a different (random) initialization of particle positions and velocities.}

\subsection{Results}
\wdf{A Fourier analysis of the velocity field is used to determine the onset of coherent vortex motion \cite{GTZ93,MGHEH12}, indicating the onset of instability. The Fourier transform of the momentum density is given by
\beq
\label{7.1}
{\bf \hat{p}} (k) = \frac{{m}}{{2\pi}} \sum\limits_{j = 1}^{{N_p}} {{ {\bf v}_j e^{i {\bf k} \cdot {\bf x}_j } }},
\eeq
where $N_p$ is the number of particles, $m$, $\bf x$ and $\bf v$ are the particle mass, position and velocity, respectively, and $\bf k$ is the wavenumber vector. Since the domain is periodic, allowed wavenumbers are ${\bf k} = \left(2 \pi a/L_x,\ 2 \pi b/L_y,\ 2 \pi c/L_z\right)$, where $a$, $b$ and $c$ are positive integers and $L_x = L_y = L_z = L$ for the current cubic system. The squared norm of the Fourier momentum density is determined by integrating Eq.\ \eqref{7.1} over concentric shells,
\beq
\label{7.2}
P(k) = \int\limits_0^{2\pi } {\int\limits_0^\pi  {\int\limits_0^{k + \delta k} {{{\left| {{\bf{\hat p}}} \right|}^2}{r^2}\sin \theta dr d\theta d\varphi }}}.
\eeq
While the particle motion remains relatively randomly distributed, $P(k)$ increases monotonically with $k$. At the onset of the velocity vortex instability however, this pattern is disturbed and $P(k)$ becomes peaked near the first mode $P_1$, i.e., $k = 2\pi / L$. Therefore, the system stability is determined by monitoring the difference between the first and second modes, $P_1 - P_2$, with a positive value indicating a low wavenumber peak in $P(k)$ corresponding to a velocity vortex instability. A system is considered to be stable only when all five replicates remain stable and, conversely, unstable when just one of the five replicates becomes unstable.}

\begin{figure}
{\includegraphics[width=0.85\columnwidth]{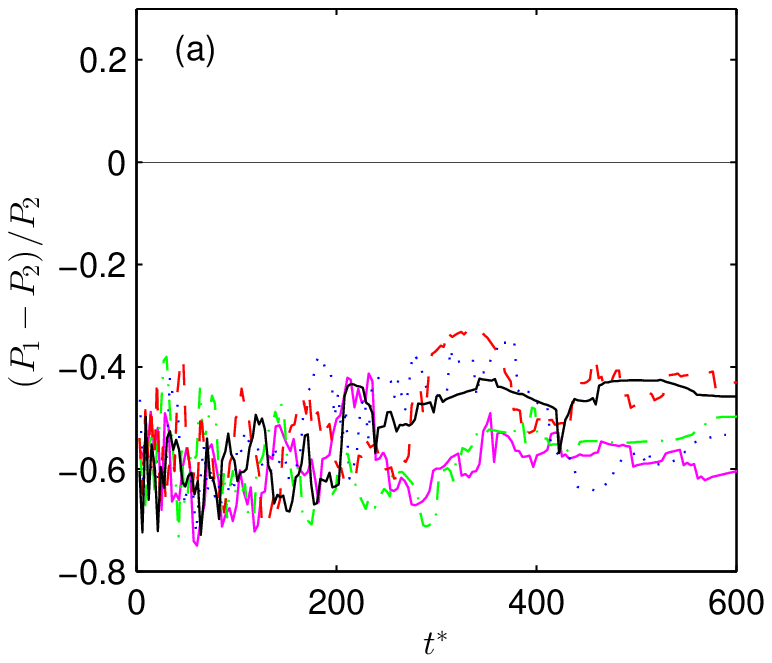}}
\\
{\includegraphics[width=0.85\columnwidth]{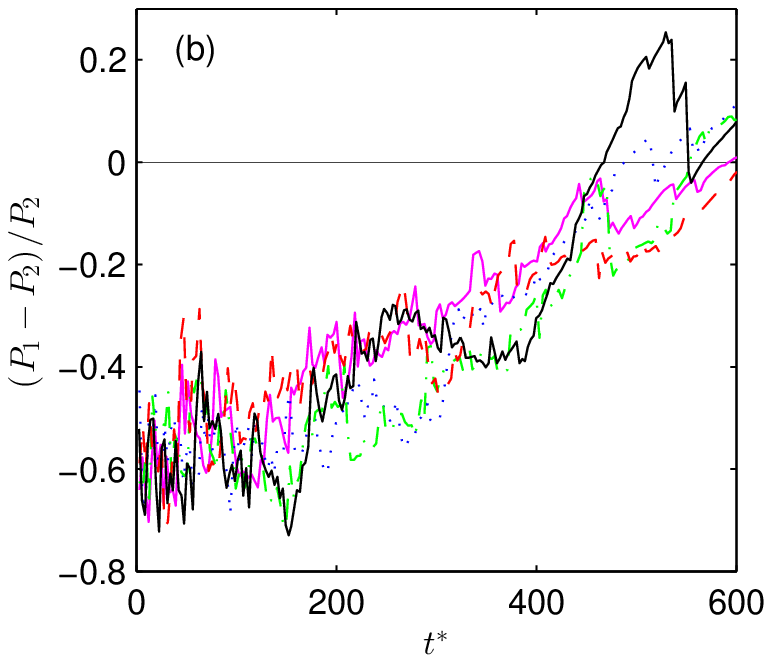}}
\caption{(Color online) Momentum modes of a stable (a) and unstable (b) cases for conditions $Re_{T_0} \approx 5$, $\alpha = 0.8$ and $\rho_s/\rho_g = 1000$ and $\phi \approx 0.3$. The unstable simulations have system size and concentration of $L^* = 4.48$ and $\phi = 0.3028$ while the smaller stable system has $L^* = 3.58$ and $\phi = 0.2843$.
\label{dns1}}
\end{figure}

\begingroup
\squeezetable
\begin{table}
\begin{center}
\caption{\wdf{Summary of principal DNS results for conditions $Re_{T_0} \approx 5$ and $\rho_s/\rho_g = 1000$.}
\label{dnsresults}}
\begin{tabular}{|c|c|c|c|c|}
\hline
$N_p$ & $\phi$ & $\alpha$ & $L^*$ & un/stable\\
\hline
131 & 0.2009  & 0.9  & 6.99  & Unstable  \\
\hline
87  & 0.2014  & 0.9  & 6.09  & Stable    \\
\hline
87  & 0.2014  & 0.8  & 6.09  & Unstable  \\
\hline
48  & 0.1989  & 0.8  & 5.02  & Stable    \\
\hline
59  & 0.3050  & 0.9  & 4.66  & Unstable  \\
\hline
46  & 0.3027  & 0.9  & 4.30  & Stable    \\
\hline
52  & 0.3028  & 0.8  & 4.48  & Unstable  \\
\hline
25  & 0.2843  & 0.8  & 3.58  & Stable    \\
\hline
\end{tabular}
\end{center}
\end{table}
\endgroup

\wdf{Figure \ref{dns1} illustrates, as examples, the momentum modes corresponding to stable and unstable conditions. Both cases shown in Figure \ref{dns1} consider the same set of conditions other than the system size, $L^*$. In panel (a), none of the five replicates achieve $P_1 \ge P_2$ and the case is considered stable. As $L^*$ is increased from 3.5 to 4.5 in panel (b),  $P_1 - P_2$ grows in time and four of the five replicates achieve the instability criterion (recall that only one is necessary to be considered unstable).} As stability analysis shows that the stability boundary is mostly sensitive to $\phi$ and $\alpha$, we chose to set $Re_{T_0} = 5$ and $\rho_s/\rho_g = 1000$, and characterized critical $L^*$ at $\phi = 0.2$ and 0.3 and $\alpha = 0.9$ and 0.8. Table \ref{dnsresults} summarizes the DNS results that are used to narrow down the stability boundaries for comparison with the linear stability analysis carried out in Sec. \ref{sec6}.

\section{Analysis: Theory vs Simulation}
\label{sec8}

\begin{figure}
\centering
\includegraphics[width=0.85\columnwidth]{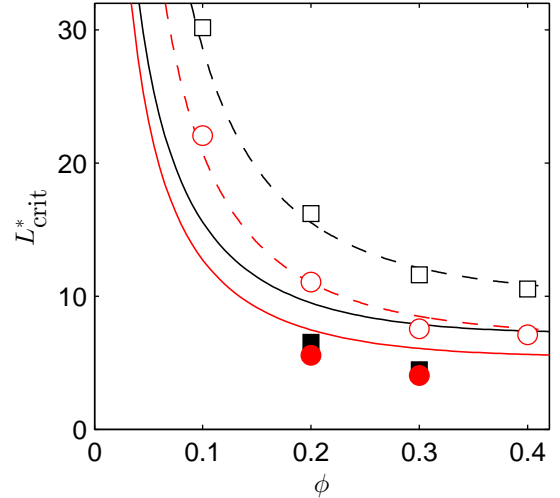}
\caption{(Color online) Neutral stability curves for the critical system size $L^*_\text{crit}$ as a function of the solid volume fraction $\phi$ for $Re_{T_0}=5$, $\rho_s/\rho_g=1000$ and two different values of the coefficient of restitution: $\alpha = 0.9$ (black lines) and $\alpha = 0.8$ (red lines). The present theory (solid lines) is compared to the dry granular theory of Ref.\ \cite{G05} (dashed lines). Results of DNS (filled symbols) and MD (empty symbols) simulations are also shown for $\alpha=0.9$ (black squares) and $\alpha=0.8$ (red circles).
\label{lsa2}}
\end{figure}

\wdf{The DNS data provided in Table \ref{dnsresults} is plotted in Fig.\ \ref{lsa2}} \vicente{as a function of the solid volume fraction $\phi$ for $Re_{T_0} \approx 5$, $\rho_s/\rho_g=1000$ and two different values of the coefficient of restitution $\alpha = 0.9$ and 0.8.} \wdf{The critical system size $L^*_\text{crit}$ is taken as the mean size of the largest stable and smallest unstable cases for each condition. Error bars are used to indicate the positions of the two cases used to calculate each $L^*_\text{crit}$, however they are obscured by the data points for clarity. Using the procedure outlined in Sec. \ref{sec6}, theoretical results for the neutral stability curves for the velocity vortex instability are also shown in Fig.\ \ref{lsa2} for the same conditions as the DNS data. In addition, the theoretical neutral stability curves corresponding to the dry granular case \cite{G05} are also compared to MD data \cite{MDCPH11} for the sake of illustration.}

It is quite apparent that qualitatively the theory captures two distinct trends that are observed in the DNS data: the inclusion of the interstitial fluid causes (i) a systematic reduction in the critical length scale and (ii) a decreased dependence on the coefficient of restitution, \wdf{i.e., inelastic dissipation}. While these trends may seem obvious simply by considering that an additional source of dissipation has been introduced, it is worthwhile to note that using a (dry) granular linear stability theory with a modified dissipation term (i.e., $\zeta_0^* \to \zeta_0^* + 2 \gamma_0^*$) does not result in a significant shift in the neutral stability curves as in \vicente{Fig.\ \ref{lsa2}}. While the current theory marks a substantial improvement over such a ``frozen'' approximation, quantitatively there are still discrepancies between the stability analysis performed here and the DNS data. On its own, this comparison may have been considered reasonable. However, the remarkable agreement of the (dry) granular theory with the MD data suggests that, perhaps, the two-fluid comparison may still be improved. One cause of the increased discrepancy could be due to the choice of the neutral stability criterion -- recall that here stability is time-dependent unlike the (dry) granular case, where neutral stability was unambiguous. On the other hand, the current results could indicate a need for improvement to the thermal drag model. A first-order extension of the thermal drag model considered here has been proposed in Ref.\ \cite{WKL03} while the most recent model proposed by Tenneti and Subramaniam \cite{TS16} is of an entirely different form \footnote{As a relatively minor extension of the present study, the thermal drag model of Ref.\ \cite{WKL03} has been considered here only to investigate its impact on the linear stability analysis results of Fig.\ \ref{lsa2}. Specifically, this extension affects the granular temperature decay rate and the hydrodynamic form of the kinetic contribution to the shear viscosity coefficient. However, this extension is first-order in $Re_T \propto T^{1/2}$, and therefore becomes independent of the granular temperature. As a result, this higher-order extension does not appreciably affect the results for $L^*_{\text{crit}}$ discussed herein.}.

\section{Summary and concluding remarks}
\label{sec9}
\wdf{
In this work, we have studied the impact of an interstitial fluid on the HCS of an assembly of inelastic, monodisperse solid particles. An instantaneous force model for the carrier fluid has been incorporated into the Enskog kinetic equation. Building on a previous work \cite{GTSH12}, the instantaneous force model is simplified here by assuming initially that the mean relative velocity between the phases is negligible, i.e., $\Delta {\bf U}\equiv {\bf U}-{\bf U}_g=\mathbf{0}$. This simplification allows for a more rigorous treatment of the thermal drag contribution of the fluid, which is the dominant fluid phase effect in terms of stability of the HCS. More specifically, here the complete granular temperature dependence of the scaled friction coefficient $\gamma^*$ (which characterizes the amplitude of the drag viscous force) is considered in the derivation of the continuum model. This feature was not accounted for in a previous derivation \cite{GTSH12} of the continuum model. As before \cite{GTSH12}, the Chapman-Enskog method is used to derive the governing equations of the suspension model and their corresponding transport coefficients. The expansion has been carried out to first-order in spatial gradients (Navier-Stokes hydrodynamic order) and the explicit forms of the (scaled) transport coefficients are given by Eqs.\ \eqref{5.5}--\eqref{5.8} where their corresponding kinetic contributions are given by Eq.\ \eqref{5.14} for the shear viscosity, Eq.\ \eqref{5.22} for the thermal conductivity and Eq.\ \eqref{5.29} for the Dufour-like coefficient.

Since the forms of the transport coefficients are at hand, a linear stability analysis is then performed on the resulting continuum theory. Unlike the (dry) granular stability analysis \cite{G05}, all four linear stability modes become functions of time through the scaled thermal drag's dependence on the (decaying) granular temperature, therefore requiring a numerical solution. A simple method to determine neutral stability is proposed and the transverse velocity mode (most unstable) is solved to study the onset of the velocity vortex instability. To assess the accuracy of the new theory and resulting linear stability analysis, a suite of DNS have also been carried out and reported herein. The theoretical predictions are able to capture the most important trends observed in the DNS data and marks a substantial improvement over the previous ``frozen'' theory \cite{GTSH12}. However, the quantitative agreement is not as favorable as previous studies have found for similar (dry) granular analyses \cite{MDCPH11}, indicating that additional refinement -- to either the underlying kinetic theory itself or the linear stability analysis -- may be possible.}

\acknowledgments

\wdf{The authors are grateful to Dr. Peter Mitrano for providing the MD data used in Fig.\ \ref{lsa2}.} \vicente{This work was initiated during a visit of W. D. F. and C. M. H. to the Departamento de F\'{\i}sica, Universidad de Extremadura and they are grateful to this institution for its hospitality and support}.  The research of V. G. has been supported by the Spanish Government through grant No. FIS2013-42840-P, partially financed by FEDER funds and by the Junta de Extremadura (Spain) through Grant No. GR15104. \wdf{W. D. F. and C. M. H. would like to acknowledge the funding provided by the National Science Foundation, Grant CBET-1236157. X. Y. would like to acknowledge the funding provided by the National Science Foundation, Grant CBET-1236490.}

\appendix
\section{First-order solution}
\label{appA}

To first order, the velocity distribution function $f^{(1)}$ obeys the kinetic equation
\beqa
\label{a0}
\left(\partial_{t}^{(1)}+{\cal L}\right)f^{(1)}&-&\frac{\gamma}{m}
\frac{\partial}{\partial{\bf V}}\cdot {\bf V} f^{(1)}=-\left(D_t^{(1)}+\mathbf{V}\cdot \nabla\right.\nonumber\\
& & \left.+{\bf g}\cdot
\frac{\partial}{\partial {\bf V}}\right)f^{(0)}+J_{\text{E}}^{(1)}[f],
\eeqa
where $D_t^{(1)}\equiv \partial_t^{(1)}+\mathbf{U}\cdot \nabla$, $J_{\text{E}}^{(1)}[f]$ denotes the first-order contribution to the expansion of the Enskog collision operator and ${\cal L}f^{(1)}=-\left(J_{\text{E}}^{(0)}[f^{(0)},f^{(1)}]+J_{\text{E}}^{(0)}[f^{(1)},f^{(0)}]\right)$. Note that gravity has been assumed to be of first order in spatial gradients in Eq.\ \eqref{a0}. The macroscopic balance equations to first order in gradients give
\begin{widetext}
\beq
\label{a0.1}
D_t^{(1)}n=-n\nabla \cdot \mathbf{U}, \quad D_t^{(1)}\mathbf{U}=-\rho^{-1}\nabla p+\mathbf{g},
\quad D_t^{(1)}T=-\frac{2p}{d n}\nabla \cdot \mathbf{U}-\zeta^{(1)} T.
\eeq
Substitution of Eq.\ \eqref{a0.1} into the right-hand side of Eq.\ \eqref{a0} allows us to get the form of the first-order distribution $f^{(1)}$. It is given by
\beq
\label{a1}
f^{(1)}=\boldsymbol{\mathcal{A}}\left(
\mathbf{V}\right)\cdot  \nabla \ln
T+\boldsymbol{\mathcal{B}}\left(
\mathbf{V}\right) \cdot \nabla \ln n
+\mathcal{C}_{ij}\left( \mathbf{V} \right)\frac{1}{2}\left( \partial _{i}U_{j}+\partial _{j
}U_{i}-\frac{2}{d}\delta _{ij}\nabla \cdot
\mathbf{U} \right)+\mathcal{D}\left( \mathbf{V} \right) \nabla \cdot
\mathbf{U},
\eeq
where the quantities $\boldsymbol{\mathcal{A}}\left(\mathbf{V}\right)$, $\boldsymbol{\mathcal{B}}\left(
\mathbf{V}\right)$, $\mathcal{C}_{ij}\left( \mathbf{V} \right)$ and $\mathcal{D}\left( \mathbf{V} \right)$ are the solutions of the following linear integral equations:
\beq
\label{a2}
-\left(\frac{2\gamma}{m} +\zeta^{(0)}\right)T\frac{\partial \boldsymbol{\mathcal{A}}}{\partial T}-\frac{1}{2}\zeta^{(0)}
\boldsymbol{\mathcal{A}}-\frac{\gamma}{m}\frac{\partial}{\partial {\bf V}}\cdot {\bf V}\boldsymbol{\mathcal{A}}
+{\cal L}\boldsymbol{\mathcal{A}}={\bf A},
\eeq
\begin{equation}
\label{a3}
-\left(\frac{2\gamma}{m} +\zeta^{(0)}\right)T\frac{\partial \boldsymbol{\mathcal{B}}}{\partial T}
-\frac{\gamma}{m}\frac{\partial}{\partial {\bf V}}\cdot {\bf V}\boldsymbol{\mathcal{B}}
+{\cal L}\boldsymbol{\mathcal{B}}
={\bf B}+\left[\frac{2n}{m}\frac{\partial \gamma}{\partial n}+\zeta ^{(0)}\left(1+\phi
\frac{\partial \ln \chi}{\partial \phi} \right)\right]\boldsymbol{\mathcal{A}},
\end{equation}
\begin{equation}
\label{a4}
-\left(\frac{2\gamma}{m} +\zeta^{(0)}\right)T\frac{\partial \mathcal{C}_{ij}}{\partial T} -\frac{\gamma}{m}\frac{\partial}{\partial {\bf V}}\cdot {\bf V}\mathcal{C}_{ij}+{\cal L}\mathcal{C}_{ij}=C_{ij},
\end{equation}
\begin{equation}
\label{a5}
-\left(\frac{2\gamma}{m} +\zeta^{(0)}\right)T\frac{\partial \mathcal{D}}{\partial T}-\frac{\gamma}{m}\frac{\partial}{\partial {\bf V}}\cdot {\bf V}\mathcal{D}+{\cal L}\mathcal{D}=D.
\end{equation}
\end{widetext}
Here, the expressions of the inhomogeneous terms $\mathbf{A}$, $\mathbf{B}$, $C_{ij}$ and $D$ are given by equations (A5)--(A8), respectively, of Ref.\ \cite{GTSH12}. It must be remarked that the integral equations \eqref{a2}--\eqref{a5} differ from those obtained in Ref.\ \cite{GTSH12} by the action of the operator $T\partial_T$ on the unknowns $\boldsymbol{\mathcal{A}}$, $\boldsymbol{\mathcal{B}}$, $\mathcal{C}_{ij}$ and $\mathcal{D}$:
\beq
\label{a6}
T \partial_T X(\mathbf{V},\gamma) =-\frac{1}{2}\frac{\partial}{\partial {\bf V}}\cdot {\bf V} X
-\frac{1}{2}\gamma \frac{\partial X}{\partial \gamma},
\end{equation}
where $X\equiv \left\{\boldsymbol{\mathcal{A}}, \boldsymbol{\mathcal{B}}, \mathcal{C}_{ij}, \mathcal{D} \right\}$. In addition, the external field does explicitly appear in the above integral equations. This is due to the particular form of the gravitational force \cite{CC70}.

The kinetic coefficients $\eta_k$, $\kappa_k$ and $\mu_k$ are defined as
\beq
\label{a7}
\eta_k=-\frac{1}{(d-1)(d+2)}\int\; \dd{\bf v}\; D_{ij}({\bf V}) \mathcal{C}_{ij}({\bf V}),
\eeq
\begin{equation}
\label{a8}
\kappa_k=-\frac{1}{dT}\int\, \dd{\bf v}\; {\bf S}({\bf V})\cdot {\boldsymbol {\mathcal A}}({\bf V}),
\end{equation}
\begin{equation}
\label{a9}
\mu_k=-\frac{1}{dn}\int\, \dd{\bf v}\; {\bf S}({\bf V})\cdot {\boldsymbol {\mathcal B}}({\bf V}),
\end{equation}
where
\begin{equation}
\label{a10}
D_{ij}({\bf V})=m(V_iV_j-\frac{1}{d}V^2\delta_{ij}),
\eeq
\beq
\label{a10bis}
{\bf S}({\bf V})=\left(\frac{m}{2}V^2-\frac{d+2}{2}T\right){\bf V}.
\end{equation}
In order to determine $\eta_k$, we multiply both sides of Eq.\ \eqref{a4}  by $D_{ij}(\mathbf{V})$ and integrate over velocity. In the case of $\kappa_k$ and $\mu_k$, we multiply Eqs.\ \eqref{a2} and \eqref{a3}, respectively, by $\mathbf{S}(\mathbf{V})$ and integrate over $\mathbf{V}$. After some algebra, one achieves the first-order differential equations \eqref{5.9}, \eqref{5.20} and \eqref{5.25} for the (dimensionless) transport coefficients $\eta_k^*$, $\kappa_k^*$ and $\mu_k^*$, respectively.

\section{Dufour-like and Euler transport coefficients}
\label{appB}

The hydrodynamic form of the kinetic contribution to the (scaled) Dufour-like transport coefficient can be written as
\beq
\label{5.29}
\mu_{k,\text{hyd}}^*=\frac{\Lambda}{\nu_\kappa^*(\zeta_0^*+2\gamma^*)^3(\zeta_0^*-2\nu_\kappa^*)(3\zeta_0^*-2\nu_\kappa^*)
(\zeta_0^*-\nu_\kappa^*)},
\eeq
where we have introduced the quantity
\beqa
\label{5.30}
\Lambda &=& \nu_\kappa^*(\zeta_0^*+2\gamma^*)\left[B \gamma^*\zeta_0^*(2\nu_\kappa^*-3\zeta_0^*)\right.\nonumber\\
& & \times\left(6\gamma^*\zeta_0^*+\zeta_0^{*2}-
8\gamma^*\nu_\kappa^*-2\zeta_0^*\nu_\kappa^*\right)-2C (2\gamma^*+\zeta_0^*)^2\nonumber\\
& & \times \left.(\zeta_0^*-2\nu_\kappa^*)(\zeta_0^*-\nu_\kappa^*)\right]
+4B\gamma^{*3}\zeta_0^*\left(1+\frac{2\gamma^*}{\zeta_0^*}\right)^{2\nu_\kappa^*/\zeta_0^*}\nonumber\\
& & \times
\left(3\zeta_0^{*3}-11\zeta_0^{*2}\nu_\kappa^*
+12\zeta_0^*\nu_\kappa^{*2}-4\nu_\kappa^{*3}\right)\nonumber\\
& & \times
{_2F_1}\left(\frac{2\nu_\kappa^*}{\zeta_0^*}, \frac{2\nu_\kappa^*}{\zeta_0^*},1+\frac{2\nu_\kappa^*}{\zeta_0^*},-\frac{2\gamma^*}{\zeta_0^*}\right).
\eeqa
In Eq.\ \eqref{5.30}, we have introduced the quantities
\beq
\label{5.31}
B\equiv 2\kappa_k^*\phi \partial_\phi \ln R_\text{diss}(\phi),
\eeq
\beqa
\label{5.32}
C &\equiv& \kappa_k^* \zeta_0^{*}\left(1+\phi\partial_{\phi}
\ln \chi \right)+\frac{d-1}{d}a_2\nonumber\\
& &+3\frac{2^{d-2}(d-1)}{d(d+2)}\phi \chi
(1+\alpha)\left(1+\frac{1}{2}\phi\partial_\phi\ln\chi\right)\nonumber\\
& & \times
\left[\alpha(\alpha-1)+\frac{a_2}{6}(10+2d-3\alpha+3\alpha^2)\right].
\eeqa

The Euler transport coefficient $\zeta_U$ is given by
\beq
\label{b2}
\zeta_U=\zeta_{10}+\zeta_{11},
\eeq
where
\begin{equation}
\label{b3}
\zeta_{10}= -3\frac{2^{d-2}}{d}\chi \phi (1-\alpha^2),
\end{equation}
and
\beqa
\zeta_{11}&=&\frac{1}{2nT}\frac{\pi ^{(d-1)/2}}{d\Gamma \left( \frac{d+3}{2} \right)}
\sigma ^{d-1}\chi m (1-\alpha^{2})\nonumber\\
& & \times \int \dd\mathbf{V}_{1}\,\int
\dd\mathbf{V}_{2}\,g_{12}^{3}\;f^{(0)}(\mathbf{V}_{1})\mathcal{D}(\mathbf{V}_{2}). \label{b4}
\eeqa
The function $\mathcal{D}(\mathbf{V})$ is the solution to the linear
integral equation (\ref{a5}). An approximate solution to Eq.\ \eqref{a5} can be obtained by
taking the leading Sonine approximation
\begin{equation}
\mathcal{D}(\mathbf{V})\rightarrow e_{D}f_{M}(\mathbf{V})F(\mathbf{V}), \label{b5}
\end{equation}
where
\begin{equation}
F(\mathbf{V})=\left(\frac{m}{2T}\right) ^{2}V^{4}-\frac{d+2}{2}\frac{m}{T}V^{2}+\frac{d(d+2)}{4},
\eeq
and
\beq
\quad f_{M}(\mathbf{V})=n \left(\frac{m}{2\pi T}\right)^{d/2}e^{-mV^2/2T}  \label{b6}
\end{equation}
is the Maxwellian distribution.  The coefficient $e_D$ is given by
\begin{equation}
e_{D}=\frac{2}{d(d+2)}\frac{1}{n}\int \;\dd\mathbf{V}\;\mathcal{D}(
\mathbf{V})F(\mathbf{V}). \label{b7}
\end{equation}
The relation between $\zeta_{11}$ and $e_D$ is
\begin{equation}
\label{b8}
\zeta_{11}=\frac{3(d+2)}{32d}\chi (1-\alpha^2)
\left(1+\frac{3}{128}a_2\right)e_D^*,
\end{equation}
where $e_D^*=\nu e_D$.  The coefficient $e_D^*$ is determined by substituting (\ref{b5}) into the integral
equation (\ref{a5}), multiplying by $F({\bf V})$ and integrating over ${\bf V}$. The
result is
\beqa
\label{b9}
& & \frac{2\gamma^*+\zeta_0^*}{2}\gamma^*\frac{\partial e_D^*}{\partial \gamma^*}+\left(\gamma^*+\nu_\gamma^*-\frac{3}{2}\zeta_0^*\right)
e_D^*=\nonumber\\
& & \frac{9}{256} \frac{2^{d} (d+2)}{d^2} \chi \phi \left[\frac{\omega^*}{2(d+2)}-\frac{1+\al}{2}\left(\frac{1}{3}-\al\right)a_2\right],
\nonumber\\
\eeqa
where
\begin{equation}
\label{b10}
\nu_\gamma^*=-\frac{1+\alpha}{192}\chi \left[30\alpha^3-30\alpha^2+(105+24 d) \alpha-56d-73\right],
\end{equation}
\beqa
\label{b11}
\omega^*&=&(1+\alpha)\left\{
(1-\alpha^2)(5\alpha-1)-\frac{a_2}{6}\left[15\alpha^3\right.\right.\nonumber\\
& & \left.\left.-3\alpha^2+3(4d+15)\alpha-(20d+1)\right]\right\}.
\eeqa
As in the case of the other kinetic coefficients, if one neglects the term $\gamma^*\partial_{\gamma^*}e_D^*$ in Eq.\ \eqref{b9}, one simply gets
\beq
\label{b12}
e_{D,\text{approx}}^*=\frac{\frac{9}{256}\frac{2^{d} (d+2)}{d^2}\chi \phi \left[\frac{\omega^*}{2(d+2)}-\frac{1+\al}{2}\left(\frac{1}{3}-\al\right)a_2\right]}
{\gamma^*+\nu_\gamma^*-\frac{3}{2}\zeta_0^*}.
\eeq
The expression \eqref{b12} was already derived in our previous work \cite{GTSH12}. The hydrodynamic solution to Eq.\ \eqref{b9} is
\beqa
\label{b13}
e_{D,\text{hyd}}^*&=&\frac{9}{128}\frac{2^{d} (d+2)}{d^2}\chi \phi \gamma^{*3}\zeta_0^*\Delta
\left(2\gamma^*+\zeta_0^*\right)^{-4}\nonumber\\
& \times &   \left[\frac{\omega^*}{2(d+2)}-\frac{1+\al}{2}\left(\frac{1}{3}-\al\right)a_2\right],
\eeqa
where
\begin{widetext}
\beqa
\label{b14}
\Delta&=&12\frac{2\gamma^*+\zeta_0^*}
{\gamma^*\left(2\nu_\gamma^*-\zeta_0^*\right)}+\frac{4}{\nu_\gamma^*}\;{_2F_1}\left(\frac{2\nu_\gamma^*}{\zeta_0^*}, \frac{2\nu_\gamma^*}{\zeta_0^*},1+\frac{2\nu_\gamma^*}{\zeta_0^*},-\frac{2\gamma^*}{\zeta_0^*}\right) +\frac{\zeta_0^{*3}}{\gamma^{*3}(2\nu_\gamma^*-3\zeta_0^*)}
\nonumber\\
& \times & {_2F_1}\left(\frac{2\nu_\gamma^*}{\zeta_0^*}, -3+\frac{2\nu_\gamma^*}{\zeta_0^*},-2+\frac{2\nu_\gamma^*}{\zeta_0^*},-\frac{2\gamma^*}{\zeta_0^*}\right)+\frac{3\zeta_0^{*2}}{\gamma^{*2}(\nu_\gamma^*-\zeta_0^*)}
\;{_2F_1}\left(\frac{2\nu_\gamma^*}{\zeta_0^*}, -2+\frac{2\nu_\gamma^*}{\zeta_0^*},-1+\frac{2\nu_\gamma^*}{\zeta_0^*},-\frac{2\gamma^*}{\zeta_0^*}\right).\nonumber\\
\eeqa
\end{widetext}

\bibliographystyle{apsrev}



\end{document}